\documentclass[twocolumn]{aastex63}
\usepackage[maxfloats=256]{morefloats}
\newcommand\clm{}

\received{November 22, 2023}
\revised{March 12, 2024}
\accepted{March 14, 2024}
\submitjournal{ApJ}

\shorttitle{Spectral Mapping of a Reionization-era Analog}
\shortauthors{Martin et al.}

\newcommand{\rc}{\mbox {R$_C$}}
\newcommand{\rx}{\mbox {R$_X$}}

\newcommand{\aFe}{\mbox {$\alpha / {\rm Fe}$}}
\newcommand{\xionhe}{\mbox {$\xi_{ion}({\rm He}^+)$}}
\newcommand{\xion}{\mbox {$\xi_{ion}$}}
\newcommand{\xionh}{\mbox {$\xi_{ion}({\rm H})$}}

\newcommand{\ho}{\mbox {${\rm H_{0}}$}}

\newcommand{\nev}{\mbox {[Ne V]}}

\newcommand{\ariv}{\mbox {[Ar IV]}}

\newcommand{\oiii}{\mbox {[O III]}}
\newcommand{\ciii}{\mbox {C III]}}

\newcommand{\feiv}{\mbox {[Fe IV]}}
\newcommand{\fev}{\mbox {[Fe V]}}

\newcommand{\oii}{\mbox {[O II]}}

\newcommand{\sii}{\mbox {[S II]}}

\newcommand{\oi}{\mbox {[O I]}}
\newcommand{\cliii}{\mbox {[Cl III]}}

\newcommand{\rhplus}{\mbox      {R$_{\rm S}$(H$^+$)}}

\newcommand{\rheplusplus}{\mbox {R$_{\rm S}$(He$^{+2}$)}}
\newcommand{\qheplus}{\mbox {Q(He$^+$)}}
\newcommand{\qh}{\mbox {Q(H)}}
\newcommand{\hplus}{\mbox {H$^+$}}
\newcommand{\heplus}{\mbox {He$^+$}}
\newcommand{\heplusplus}{\mbox {He$^{+2}$}}
\newcommand{\heii}{\mbox {He II}}
\newcommand{\hei}{\mbox {He I}}
\newcommand{\Ha}{\mbox {H$\alpha$}}
\newcommand{\Hb}{\mbox {H$\beta$}}

\newcommand{\hi}{\mbox {H I}}
\newcommand{\hii}{\mbox {H II}}

\newcommand{\fesc}{\mbox {$f_{esc}$}}
\newcommand{\fescHe}{\mbox {$f_{esc,54}$}}  
\newcommand{\lya}{\mbox {Ly$\alpha$}}

\newcommand{\fv}{\mbox {$\epsilon$}}  
\newcommand{\fvh}{\mbox {$\epsilon_h$}}  
\newcommand{\nmsq}{\mbox {$\langle n_e^2 \rangle$}}  
\newcommand{\nrmsq}{\mbox {$\langle n_e^2 \rangle^{1/2}$}}  

\newcommand{\kms}{\mbox {$ {\rm km~s}^{-1}$}}

\newcommand{\msun}{\mbox {$ {\rm ~M_\odot}$}}
\newcommand{\zsun}{\mbox {$ {\rm ~Z_\odot}$}}

\newcommand{\msunyr}{\mbox {$ {\rm ~M_\odot}$~yr$^{-1}$}}
\newcommand{\Mr}{\mbox {$ {\rm ~M_r}$}}

\newcommand{\ergsec}{\mbox {~ergs~s$^{-1}$}}
\newcommand{\flux}{\mbox   {ergs~s$^{-1}$~cm$^{-2}$}}

\begin{document}

\title{Resolving the Mechanical and Radiative Feedback in
J1044+0353 with KCWI Spectral Mapping}

\correspondingauthor{Crystal Martin}
\email{cmartin@ucsb.edu}

\author[0000-0001-9189-7818]{Crystal L. Martin}
\affiliation{Department of Physics \\
University of California Santa Barbara \\ 
Santa Barbara, CA 93106, USA }

\author{Zixuan Peng}
\affiliation{Department of Physics \\ 
University of California Santa Barbara \\
 Santa Barbara, CA 93106, USA }

\author{Yuan Li}
\affiliation{Department of Physics \\ 
University of California Santa Barbara \\
 Santa Barbara, CA 93106, USA }



\begin{abstract}
We present integral field spectroscopy toward and around J1044+0353, a rapidly 
growing, low-metallicity galaxy which produces extreme \oiii\ line emission. A
new map of the O32 flux ratio reveals a density-bounded ionization 
cone emerging from the starburst. 
{\clm 
The interaction of the hydrogen ionizing radiation, produced by the 
very young starburst, with a cavity previously carved out by a galactic outflow, 
whose apex lies well outside the starburst region, determines the pathway for 
global Lyman continuum (LyC) escape.
}
In the region within a few hundred parsecs of the young starburst, we demonstrate
that superbubble breakthrough and blowout contribute distinct components to the \oiii\
line profile, broad and very-broad emission line wings, respectively. We draw
attention to the large \oiii\ luminosity of the broad component and argue that
this emission comes from photoionized, superbubble shells rather than a galactic
wind as is often assumed.
{\clm
The spatially resolved \heii\ $\lambda 4686$ nebula appears to be photoionized
by young star clusters. Stellar wind emission from these stars is likely the
source of line wings detected on the \heii\ line profile.
This broader \heii\ component indicates slow stellar
winds, consistent with an increase in stellar rotation (and a decrease
in effective escape speed) at the metallicity of J1044+0353.
}
At least in J1044+0353, the recent star formation history plays a critical 
role in generating a global pathway for LyC escape, and the anisotropic escape 
would likely be missed by direct observations of the LyC. 
\end{abstract}

%


\section{Introduction} \label{sec:intro}

{\clm
Most of the hydrogen in the universe underwent a phase change, from neutral to ionized,
during the first billion years of galaxy assembly.  Stars more massive than roughly 20\msun\ 
produce Lyman continuum (LyC) photons at a significant rate, so LyC leakage from star-forming 
galaxies is widely believed to be the dominant source of the radiation that ionized the 
intergalactic gas \citep{Bouwens2015b,Finkelstein2019}. How this LyC radiation leaked out of 
galaxies, however, is still not well understood.  Integral field spectroscopy of the nearest analogs 
of the galaxies that likely drove cosmic reionization can provide critical insight into 
how reionization proceeded.
}

{\clm
At redshifts where the LyC can be directly observed, neither galaxy metallicity nor stellar 
mass provide a reliable prediction of the LyC escape fraction, \fesc.  Two of the best indicators
of high \fesc\ are a high star formation rate (SFR) surface density or a high nebular \oiii\ 
to \oii\ flux ratio \citep{Flury2022}. High \oiii\ / \oii\ ratios, for example, are consistent 
with density-bounded channels, where the column density of neutral hydrogen is insufficient 
to absorb all the LyC radiation \citep{Jaskot2013,Izotov2018b,Plat2019,Ramambason2020}. Similarly, 
supernova explosions may be more likely to open pathways for LyC escape from starbursts with 
high SFR surface density since their measured outflow speeds increase with SFR surface density
\citep{Kornei2012,Heckman2016}. Yet efforts to demonstrate correlations between these quantities
and \fesc\ yield confusing results: (1)  outflow velocities,  measured from the Doppler shifts 
of interstellar absorption lines, have not shown
a correlation with \fesc\ \citep{Henry2015,Jaskot2017},  and (2) density-bounded channels are not 
a unique explanation for high \oiii\ / \oii\ ratios \citep{Plat2019,Flury2022}. 
}
The resulting tension has been cited as evidence that winds might become ineffective at the highest SFR 
surface densities due to catastrophic radiative cooling \citep{Gray2019b,Danehkar2021}. It is 
therefore important to understand whether line-of-sight effects caused by anisotropic feedback 
might offer an alternative explanation for these apparent contradictions.

We observed J1044+0353 with multiple pointings of the Keck Cosmic Web Imager (KCWI), 
an integral field spectrograph on the Keck~II telescope.  The properties of our target 
have many similarities to reionization-era galaxies and are listed in Table~\ref{tab:basic_properties}. 
The \oiii\ equivalent width is {\clm 1400} \AA\  in  the  $\lambda 5007$ line alone, an extreme 
value  at any redshift and an indication of efficient ionizing photon production \citep{Chevallard2018}.
The high specific star formation rate,  $4.1 \times 10^{-8}$~yr$^{-1}$ \citep{Berg2022}, underlines
this small galaxy's rapid, recent growth.
The far-ultraviolet (FUV) spectrum of J1044+0353 shows strong, high-ionization emission lines, including
\heii\ $\lambda 1640$ \citep{Berg2016,Berg2022}. Recombination of He$^{+2}$ produces \heii\ $\lambda 1640$ 
and optical \heii\ $\lambda 4686$ emission but requires photons with energies $ \ge\ 4$ Rydberg. Normal stellar 
populations, as described by Binary Population and Spectral Synthesis v2.1 (BPASS) models \citep{Eldridge2017} 
for example, do not produce enough 54.4~eV photons to explain the \heii\ $\lambda 1640$ emission from J1044+0353
\citep{Berg2019,Berg2021,Olivier2022}. This high-energy ionizing photon production problem has been observed in 
other local galaxies \citep{Shirazi2012,Jaskot2013,Stasinska2015,Senchyna2017,Schaerer2019,Mingozzi2022,Mingozzi2023}
and is a common feature reionization-era galaxies \citep{Stark2015b,Mainali2017,Bunker2023,Senchyna2023,Topping2024}. 
The damped \lya\ profile in the same FUV spectrum of J1044+0353 \citep{Hu2023}, however, indicates a low \fesc\ 
value along our sightline. 
{\clm
In known LyC leakers, the intensity of \heii\ relative to \Hb\ appears to be driven by metallicity rather than 
\fesc\ \citep{Marques-Chaves2022}. J1044+0353 has a gas-phase oxygen abundance of $12 + \log(O/H) = 7.45$
\citep{Peng2023} but lacks the strong nitrogen enhancement seen  in GN-z11  \citep{Senchyna2023}. 
Massive stars of similar metallicity are expected to have a harder
ionizing spectrum on empirical grounds; see Figures 2-4 of \citet{Sander2022} for example.
}
Green Pea galaxies are a well-studied sample selected based on their high \oiii\ equivalent 
widths  \citep{Cardamone2009}, but their redshifts, $0.112 < z <  0.360$,  place them over ten times further 
away than J1044+0353. 

\begin{figure}[h]
\includegraphics[angle=0,height=3.25in, trim = 20 0 0 0]{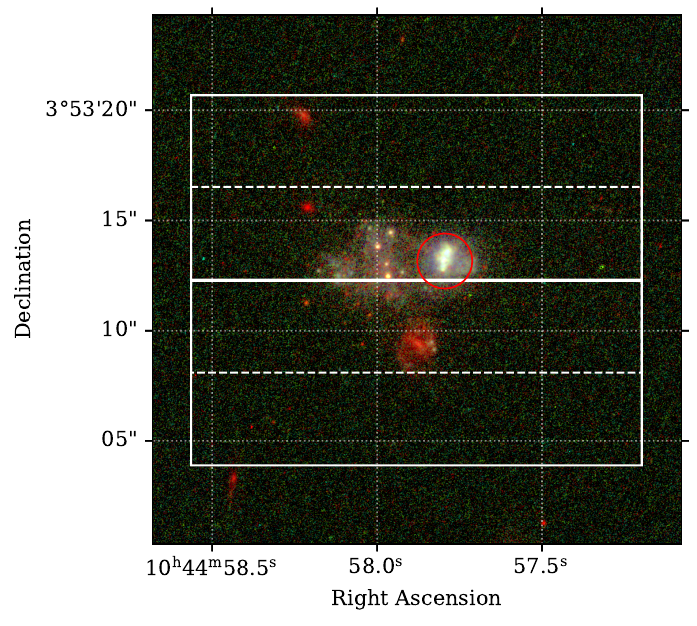}

\caption{
Composite {\it Hubble} image of J1044+0353 constructed 
from 8 filters:  F125LP, F150LP, F165LP, F336W, F438W, F606W, F665N, and F814W.
Using the pivot wavelengths, filters were mapped into the optical range. 
The stretching method is adapted from \cite{Lupton2004} but generalized to
more than three filters.
The red circle
of diameter 2\farcs5 ($\approx 350$ pc) identifies the {\it Hubble} COS
aperture used to study the starburst previously \citep{Berg2022,Xu2022,Hu2023}. 
The three slicer pointings (white) outline the footprint of the KCWI mosaic cube.
We follow the naming convention {\clm from Figure 1 of \citet{Peng2023}}:
SB for the compact starburst at 10:44:57.797, +3:53:13.16, FEN for the group 
of star clusters {\clm 1.24 kpc}
to the east 10:44:58.092, +03:53:12.58, 
and MEN and NEN for the complexes between them at 
10:44:57.964, +03:53:12.40 and 10:44:57.986, +03:53:13.81.
The MEN and NEN regions are resolved into 
reddish-colored clusters strung out along a north-south direction between
SB and FEN. The Balmer absorption is strong across MEN, NEN, and FEN, {\clm so we refer
to them as the post-starburst region.}
The red spiral galaxy visible south of J1044+0353 at
10:44:57.874, +03:53:09.39
has an emission-line redshift of $z = 0.27522$.
}
\label{fig:kcwi_fov_duo} \end{figure}

Figure~\ref{fig:kcwi_fov_duo} illustrates the information gained from spatially resolving the starburst
environment. The compact starburst is off center in J1044+0353. In a preliminary 
analysis of our KCWI data cube \citep{Peng2023}, we mapped the star formation history across the surrounding
galaxy and discovered a post-starburst population roughly
1.3 kpc east of the starburst. Whereas the interstellar absorption lines in the FUV 
spectrum had indicated a very weak outflow \citep{Xu2022}, the global gas kinematics revealed a large, 
bipolar outflow \citep{Peng2023}. Initially, it came as a surprise that the apex of this galactic 
wind was outside the FUV spectral aperture, well separated from the compact starburst circled in
Figure~\ref{fig:kcwi_fov_duo}. However, the post-starburst population, where the stars are at least 
10-20 Myr old, has produced more supernovae than the compact starburst, which is less than 4 Myr old  
\citep{Olivier2022}. This spatial mapping of the star formation history and gas kinematics can help
us understand LyC escape.

A single, short burst of star formation fails to produce much LyC escape.
The main sequence lifetime of the lowest mass star undergoing core-collapse, $\approx 8-10$ \msun, 
is roughly an order of magnitude longer than the lifetime of the lowest mass star producing 
significant LyC radiation \citep{Leitherer1999}. Considering the steep slope of the stellar 
initial mass function (IMF), a short burst stops producing LyC radiation before most of 
the supernova feedback has taken place, a situation we call the {\it supernova timing problem}.
{\clm
Cosmological, zoom-in simulations produce channels for Lyman escape several ways,
including runaway OB stars \citep{Kimm2014}, radiative feedback \citep{Kakiichi2021,Chevance2022},
and frequent galaxy mergers in the reionization era \citep{Martin-Alvarez2023,Witten2024}.  After reionization,
the environment of dwarf galaxies evolves, yet feedback continues to play a major role in 
their evolution \citep{Agertz2020,Rey2022}. 
}
Most local, dwarf galaxies are very diffuse, inactive systems \citep{Weisz2012,Guo2016}, so 
the high specific SFR in J1044+0353 indicates that some mechanism recently drove gas inwards. 
The low gas-phase metallicity across J1044+0353, relative to the mass - metallicity 
relation, provides indirect evidence for this recent gas inflow \citep{Peng2023}.

In this paper we map the ionization structure across the field shown in Figure~\ref{fig:kcwi_fov_duo} 
in order to better understand whether the supernova feedback produced by J1044+0353's recent star 
formation history creates low-density pathways through the interstellar medium. To gain additional 
insight into the origin of the hard ionizing radiation, we compare the structure of the \heii\ nebula 
in J1044+0353 to the distribution of young star clusters visible in {\it Hubble} images and the gas 
kinematics in the starburst region. 
Our presentation is organized as follows.  In Section~\ref{sec:observations}, we fully describe the observations 
and data reduction, including the generation of the KCWI mosaic in Figure~\ref{sec:kcwi} and the 
archival {\it Hubble} imaging in \ref{sec:hubble_images}.
Section~\ref{sec:results} presents measurements of the hydrogen and helium recombination
lines; we resolve the \heplusplus\ nebula and measure total recombination-line luminosities.
Section~\ref{sec:structure} 
maps out spatial variations in the gas density and ionization structure,
and we illustrate the LyC escape pathways using the \oiii\ / \oii\ flux ratio. 
Section~\ref{sec:gas_kinematics} describes the mechanical feedback in the starburst region
and identifies two sources of high-velocity, emission-line luminosity.
Section~\ref{sec:discussion} discusses whether stellar populations modeled with a standard 
IMF can match the  measured  ionizing photon efficiency and inferred supernova power.
Section~\ref{sec:summary} summarizes the results and  their implications for 
stellar feedback and \fesc\ in the reionization era.

Throughout this paper, we correct for internal reddening
using the  ${\rm A_V} = 0.20 \pm 0.10$,
measured (from the Balmer decrement) in the starburst region 
\citep{Peng2023}, and an SMC extinction curve \citep{Gordon2003}. 
 We correct fluxes for Galactic extinction \citep{Fitzpatrick1999}
using the the dust map of \citet{Schlafly2011}, which we  accessed via the IRSA interface\footnote{
             {\url https://irsa.ipac.caltech.edu/cgi-bin/bgTools/nph-bgExec}}.
All magnitudes are in the AB system, and all emission-line equivalent widths 
are quoted in the rest frame. We adopt the
total recombination rate coefficients from \cite{Storey1995} and  
the effective  recombination rate coefficients from \cite{Pequignot1991}, both
at a fiducial electron temperature of $2.0 \times 10^4$~K \citep{Peng2023}.
The optical emission lines in the integrated KCWI spectrum indicate a spectroscopic 
redshift $z_{spec} = 0.012873$ , consistent with SDSS and COS spectra \citep{Berg2022}.
However, a model of the local velocity field places J1044+0353 further away, at
an effective redshift ${\sc ZDIST} = 0.01363\pm 0.00068$ \citep{Willick1997}. 
We adopt a flat $\Lambda {\rm  CDM}$ cosmology 
with $\Omega_\Lambda = 0.7$, $\Omega_m=0.3$, and \ho = 70 km~s$^{-1}$~Mpc$^{-1}$.  
The luminosity distance is then 59.0 Mpc, and the angular scale is 278 pc/\arcsec.

\section{Data} \label{sec:observations}

A decade before the launch of JWST, hundreds of reionization-era galaxies were identified by 
their high \oiii\ equivalent width, which produces a recognizable  broad-band infrared color 
\citep{Schaerer2010,Labbe2013,Roberts-Borsani2016,Smit2015}. JWST spectroscopy has now confirmed 
this rapid increase in \oiii\ emission line strength with redshift \citep{Matthee2023,Oesch2023}.

We selected our target, J1044+0353, from the NASA-Sloan Atlas (NSA). The NSA v0\_1.2 catalog
includes nearly 27,000 dwarf galaxies, defined here as galaxies fainter than $\Mr = -18$ 
and with NSA stellar masses less than $10^9$\msun. We searched this sample for
rest-frame \oiii\ $\lambda 5007$ equivalent width exceeding 1000 \AA\ because
strong  \oiii\ emission is a common spectral feature of high-redshift galaxies
\citep{Labbe2013,Smit2014,Smit2015,Stark2015b,Roberts-Borsani2016,
Stark2017, Laporte2017,Hashimoto2018,Berg2018,Tamura2019}.  
{\clm
We found only 35 dwarf galaxies with ${\rm W}(\oiii\ \lambda 5007)  > 1000$ \AA, 
hereafter the O3EWGT1000 sample. } Dwarf galaxies with extreme \oiii\ equivalent
width are therefore  rare at low redshift, but the O3EWGT1000 sample  includes a 
high fraction of {\clm narrow-line} \heii-emitters, 17 of the 35 galaxies including
J1044+0353. Considering that 
only 71 of the 27,000 NSA dwarf galaxy spectra are listed as \heii\ emitters 
\citep{Shirazi2012}, galaxies in the O3EWGT1000 sample are 200 times more likely than 
average to emit nebular \heii\ $\lambda 4686$.

The presence of \heii\ recombination lines requires an ionizing spectrum with a significant 
photon luminosity at an energy of 54.4 eV.  
We are observing this galaxy population with integral field spectroscopy in order to gain 
new insight about the source of the hard radiation, the escape channels for ionzing radiation, 
and the mechanical feedback. We present a case study of  J1044+0353 because our observations 
spatially resolve the very-high ionization zone, and {\it Hubble} imaging provides new insight 
the number of ionzing sources.

\subsection{Integral Field Spectroscopy with KCWI} \label{sec:kcwi}

The {\it Keck Cosmic Web Imager} (KCWI) is the blue channel of an
integral field spectrograph on the Keck II telescope \citep{Morrissey2018}.  
We observed J1044+0353 (RA=10:44:57.80, DEC=+03:53:13.2)
with KCWI on 2018 January  14 under clear skies
and average seeing of 0\farcs9. These observations were made before
commissioning was fully completed, and keywords in the FITS headers needed to be
manually synchronized with the observer's log.

We configured KCWI with the small slicer, the BL grating, 
and a camera angle that produced a central wavelength of 4600 \AA.  
In this configuration, the slits are 0\farcs35 wide,
and the 24 slices subtend an 8\farcs4 $\times$ 20\farcs4 field of view. The dichroic limited
spectral coverage to wavelengths less than  5600 \AA. Falling sensitivity shortward of 3700 
\AA\ produced a steep decline in sensitivity toward the atmospheric 
cutoff.
The combination of narrow slits plus the BL grating provided an average resolution 
$R \approx 3600$ over a bandpass which includes: (1) recombination lines from \heii, \hei, and 
the hydrogen Balmer series, and (2) collisionally excited metal lines from ions 
spanning a wide range of ionization states.

We obtained exposures of 1200~s and 300~s at three KCWI pointings. As illustrated in 
Figure~\ref{fig:kcwi_fov_duo}, the slit position angle was 90 degrees east of north, and 
offsets were made in the direction perpendicular to the slits. The first exposure was 
centered on the compact starburst, and we then offset roughly half the slicer width north 
and south, thereby covering a 20\farcs4 by 16\arcsec\ field of view. We made the offsets in 
half-slice multiples to improve the point spread function (PSF) sampling. This strategy partially offsets the coarse 
sampling of the slits relative to the finer pixel sampling along the slit, 0\farcs146 pixel$^{-1}$, 
providing nearly uniform sampling in both spatial dimensions. To measure the sky spectrum,  
the slicer was chopped to a blank field, and 600~s sky exposures were made between pairs
of 1200~s and 300~s on target exposures.  This sequence provided 4500~s of integration time 
over the central half of the final mosaic. These exposures detected the bright emission 
from \oiii\ $\lambda \lambda 4959, 5007$ over the entire field of view.

\subsubsection{Cube Construction}

We used the KCWI Data Extraction and Reduction Pipeline v1.1.0 \citep[KDERP]{Morrissey2018} to 
reduce the raw frames and build calibrated data cubes. We first produced bias subtracted and
gain-corrected intensity frames with cosmic rays and scattered light subtracted. We then
mapped each detector pixel to a slice, position, and air wavelength. We extracted 
the sky spectrum from the paired sky observations, modeled the 2D sky spectrum for each integration,
and subtracted the sky counts prior to cube generation.  Each cube was corrected for differential 
atmospheric refraction based on the airmass of the observation, the position angle of 
the slicer, and the parallactic angle.  Each stage of the pipeline produced a matching variance image
or cube that describes the detector noise and the shot noise from photon counting statistics.

We observed the spectrophometric standard star Feige 34 with the same instrument configuration
during the observing run. We extracted the Feige 34 stellar spectrum, subtracted the sky 
spectrum recorded on the star-free region of the slicer, and corrected the spectrum for atmospheric 
attenuation  \citep{Buton2013} according to the airmass of the observation. 
We computed the inverse sensitivity in narrow bands that excluded
strong stellar absorption lines. We fit the inverse sensitivity with a high-order polynomial across the 
$\lambda 3650$ to $\lambda 5625$ bandpass. The extracted, flux-calibrated spectrum of Feige 34
is consistent with its reference spectrum to  within $\pm 3 \% $ across this bandpass. We applied
this inverse sensitivity function to each cube. The measurements reported in this paper
use this absolute flux calibration.\footnote{ 
       The steepening of the inverse sensitivity function shortward of  $\lambda 3650$ was
       not easily fit, as it tended to introduce high-order wiggles at longer wavelengths. 
       Inspection of the spectrum around \nev\ $\lambda 3425$ did not reveal an emission line,
       so we did not flux calibrate the shortest wavelengths.}

\subsubsection{Construction of the Mosaic Cube}

After cropping the KCWI cubes to the valid data region, we registered the world coordinates of
the compact starburst with its position in the SDSS g-band image. For the offset frames that only 
covered half of the compact starburst, we used the telescope offsets to register them 
relative to the base pointing.
Following \cite{OSullivan2020}, we coadded the individual cubes 
using the CWITools Python 
package\footnote{https://github.com/dbosul/cwitools}.
This procedure produces a mosaic cube with 0\farcs146 $\times$ 0\farcs146 $\times$ 0.5 \AA\ 
voxels. 

A similar procedure was applied to the variance frames using the {\it -vardata True} flag.
We compared the global noise properties of the intensity cube to the variance cube 
and found the former was typically higher. 
We rescaled our variance cubes by their ratio, a factor  $\approx 1.5 $. 
The resampling  to a regular grid introduces covariance among adjacent voxels of the data
cube. We ignore this covariance in our analysis because its effect on spatial scales
larger than the point-spread-function is negligible.

A few spaxels (out of over 15,000 in the entire mosaic) have saturated \oiii\ lines. 
We identified saturated pixels with more than 60,000 counts in the raw 2D
frames. In the 1200~s exposures, saturation clips the cores of the \oiii\ $\lambda 
\lambda 4959, 5007$  and \Hb\ emission lines in slits that intersect the spatially unresolved
core of the starburst. No slits show saturated \oiii\ $\lambda 4959$ or \Hb\ emission
in the short 300~s exposures.  Whenever a measurement requires integration over a masked 
spaxel, we simply measure the \Hb\ and \oiii\ $\lambda 4959$ line flux from the matched
mosaic of 300~s cubes. 
Measurements of the \oiii\ $\lambda 5007$ line in the 300~s mosaic are 
assigned a flux equal to three times the \oiii\ $\lambda 4959$ flux, consistent with 
the respective radiative decay rates of the two transitions.

\subsubsection{Spectral Extraction}

\begin{figure}[h]
\includegraphics[angle=-90,width=3.5in]{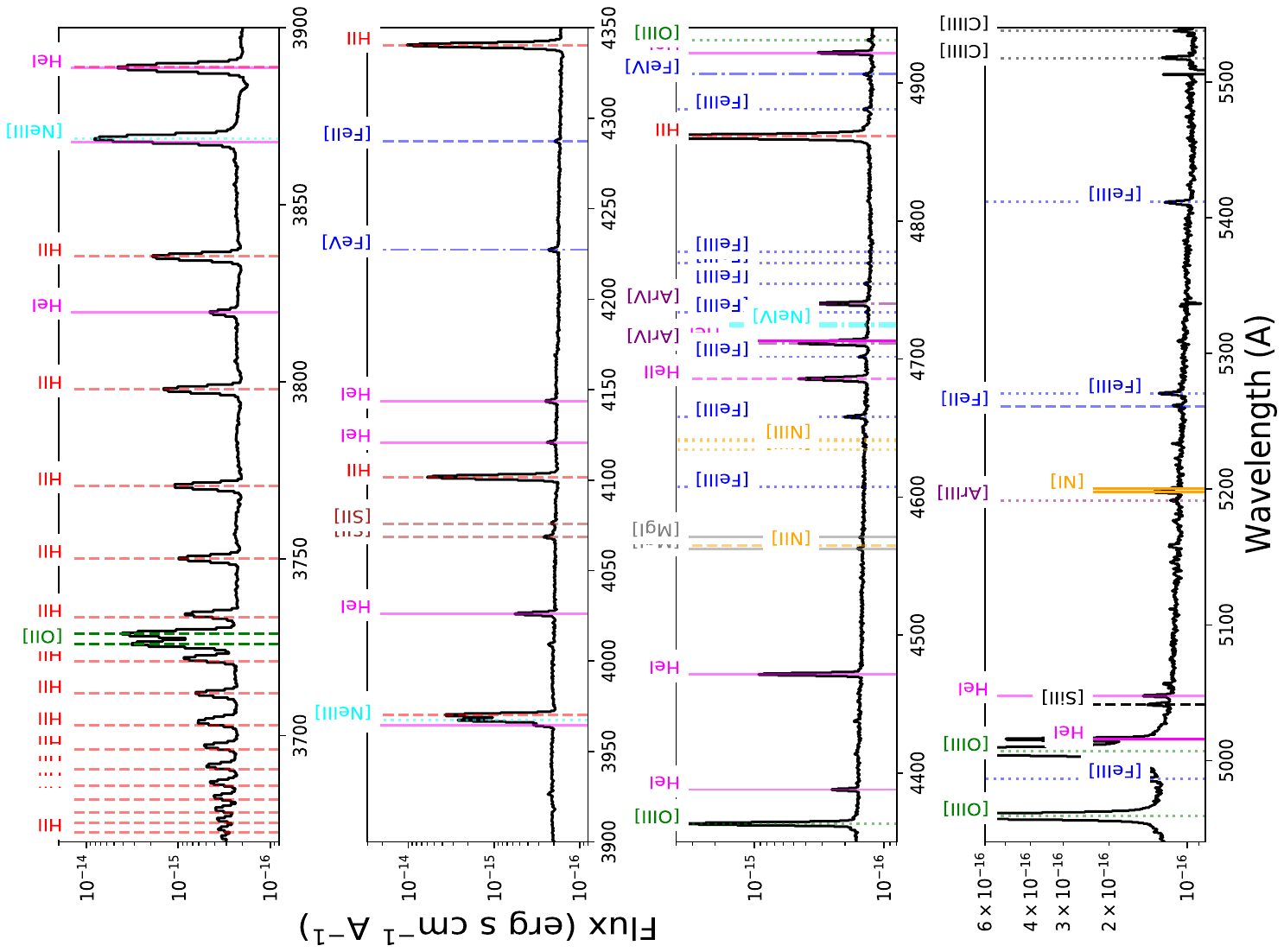}
\caption{Example rest-frame KCWI spectrum extracted from
a 3\farcs0 diameter aperture centered on the compact starburst. 
Note the prominent emission lines from the very-high ionization
zone: \heii\ $\lambda 4686$, \fev\ $\lambda 4227.4$, \feiv\ $\lambda 4906.6$, and 
\ariv\ $\lambda \lambda 4711.35, 4740.20$.
The spectral resolution separates the \oii\ $\lambda \lambda 3726.03, 28.82$
doublet lines. 
Broad, low intensity wings are visible on the \oiii\ and \Hb\ lines.
}
\label{fig:kcwi_spec} \end{figure}

The starburst spectrum shown in Figure~\ref{fig:kcwi_spec} was extracted by summing  1-D spectra within
10 spaxels of the brightest spaxel. The radius of this roughly circular aperture is 1\farcs53, similar
to the size of an SDSS fiber. We confirm the strong emission lines from low, medium, high, and very-high 
ionization line described previously by \citep{Berg2021}. 
As an external check on the flux calibration, we compared the fluxed KCWI and SDSS spectra. 
We extracted spectra at all spaxels within 
1\farcs53 of the {\sc SDSS\_PLUG} position and added them to provide a KCWI spectrum of the 
starburst.  We corrected these spectra for foreground Galactic reddening using the visual
extinction from \citet{SF2011} and the \citet{Fitzpatrick1999} average extinction curve.
To register the wavelength scales, we corrected the KCWI spectra to the heliocentric reference 
frame, and we converted the vacuum SDSS wavelengths to air.  We found that the SDSS spectrum downloaded
from the SkyServer had a higher overall flux density than our aperture-matched KCWI spectrum. Further
investigation revealed that the absolute fluxes of SDSS spectra are rather uncertain, and the
spectra should be scaled to the fiber magnitudes following \citet{Brinchmann2004}.
We multiplied the SDSS spectrum by the g, r, and i  throughput curves, and averaged the three scale
factors to obtain a multiplicative correction of 0.74. This produced a good overall agreement in
flux density between the SDSS and KCWI spectra at the red end of our bandpass. \citet{Tremonti2004} 
reported that SDSS spectra are systematically 12\% bluer than SDSS  photometry, and our comparison
confirms a discrepancy of similar magnitude in the part of the u-band covered by the SDSS spectrum.
Therefore,  the external comparison to SDSS did not add information beyond the internal calibration.

We measured W(\oiii) across a set of circular apertures centered on the starburst. In a circular
aperture of radius 1\farcs5, we measure a rest-frame W($\lambda 5007$) of nearly 1400 \AA. The 
measured EW grows rapdily with increasing aperture radius out to a radius of 2\farcs2,
where it reachs 1780 \AA, and then slowly  rises to 1930 \AA\ within an aperture of radius of 7\farcs2.
Observing J1044+0353 with a large aperture that subtends several
physical kiloparsecs produces a larger EW than does an SDSS fiber spectrum, which subtends a diameter of 830~pc. 
We conclude that observations of a J1044+0353 clone at high-redshift, where the angular diameter distance produces
scales of 5 to 8 kpc per arcsecond, through a seeing-limited 1\farcs2 slitlet (6-10~kpc in width) would 
capture most of the nebular emission. However,  spectra obtained through narrow 0\farcs2 NIRSpec shutters 
(1 to 1.6 kpc) would yield lower equivalent widths and miss much of the nebular emission.

\subsubsection{Extraction of Emission-Line Images}

We extracted pure emission-line images from the data cube. 
For each emission-line of interest, 
we defined a narrowband filter with a top-hat transmission function 
and added the continuum-subtracted line flux over the included wavelength slices.
Each synthesized filter was centered on the observed-frame wavelength 
of the line, and the filter widths ranged from 6.5 to 14.5 \AA.  
The continuum under the line was modeled using the median level measured
within two line-free windows, one on each side of the emission line. 
The continuum uncertainty was set equal to the difference between the two medians
and propagated through to the narrowband variance image.

We adaptively binned the line images to improve sensitivity to low-surface brightness features.
The narrow-band images presented a special challenge because some of the pixels had negative 
values, a result of background and continuum subtraction.  Our approach combined the robust 
treatment of negative values developed for binning background-subtracted X-ray data \citep{Sanders2001}
and the flexible bin sizes of the \cite{Cappellari2003} algorithm. Using a weighted Voronoi 
tessellation and following \citep{Diehl2006}, we adaptively binned the flux to a local signal-to-noise 
requirement.  This technique preserves spatial resolution in high S/N ratio regions while improving 
sensitivity in low surface brightness regions at the cost of reduced spatial resolution. 
Structures visible in the binned images motivated the definitions of the apertures used for
photometry, but the photometry was performed on unbinned images, avoiding the need to model
the covariance introduced by the adaptive binning.

\begin{figure}[h]
\includegraphics[angle=0,height=2.5in]{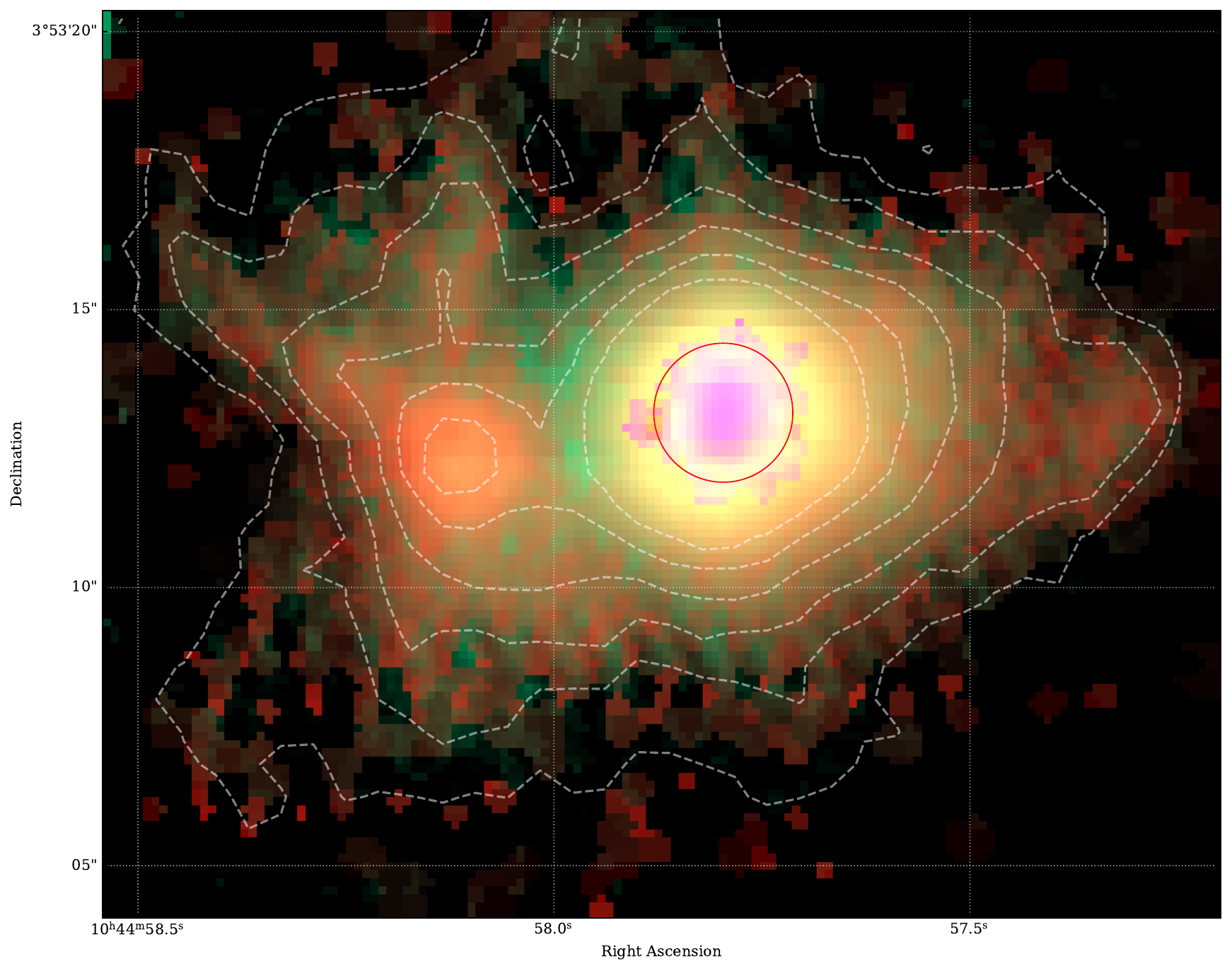}
\caption{
KCWI emission-line image of J1044+0353:  red (\Hb), green (\oiii\ $\lambda 5007$), 
violet (He II $\lambda 4686$). For scale, the red circle 
has a diameter of 2\farcs5 (695 pc). The \Hb\ to \oiii\ ratio is highest along the 
northeast and southeast lobes of the global, bipolar outflow (see \cite{Peng2023}
Figure 4). The \oiii\ nebula is spectroscopically detected across the entire field;
the \oiii\ $\lambda 5007$ contour levels are spaced by actors of two from
$1.5 \times 10^{-17}$ to $4.80 \times 10^{-16}$\flux
}

\label{fig:3color_kcwi} \end{figure}

Figure~\ref{fig:3color_kcwi} illustrates the ionization structure of the galaxy.
The very-high ionization zone emitting \heii\ $\lambda 4686$, shown in violet, is spatially resolved
but localized around the young starburst. The high-ionization gas traced by the \oiii\ emission
is detected into the circumgalactic medium, i.e., well beyond the faintest countours 
of continuum emission in the {\it Hubble}, SDSS, or DESI images.
Throughout the region east of the starburst, the Balmer emission lines sit in broad 
Balmer absorption troughs. We fit the absorption with a 
second component (having a Lorentzian profile) and extracted both a pure emission-line image, i.e. the flux
relative to the bottom of the trough, and the absorption equivalent width.

\subsection{High-resolution Morphology from {\it Hubble} Imaging} \label{sec:hubble_images}

We retrieved  {\it Hubble} images of J1044+0353 from the Mikulski Archive for Space 
Telescopes (MAST, PID 16209 \& PID 16763). The composite image in Figure~\ref{fig:kcwi_fov_duo} 
includes filters from the UV {\clm (F125LP, F150LP, F165LP)}, UV/optical (F336W,
F438W, FR462N), and red/infrared (F606W, F665N, and F814W).
We corrected the WFC3/UV frames for charge transfer effects.
Cosmic rays were removed from all frames.
Frames in each filter were drizzled onto matched images with a pixel 
scale of 0\farcs02 per pixel. We aligned each image with a g-band image from 
DESI Legacy Imaging Survey \citep{Dey2019}.
The COS and SDSS spectra of J1044+0353 cover only the starburst region.

\begin{figure}[h!]
\includegraphics[angle=90,height=5.0in]{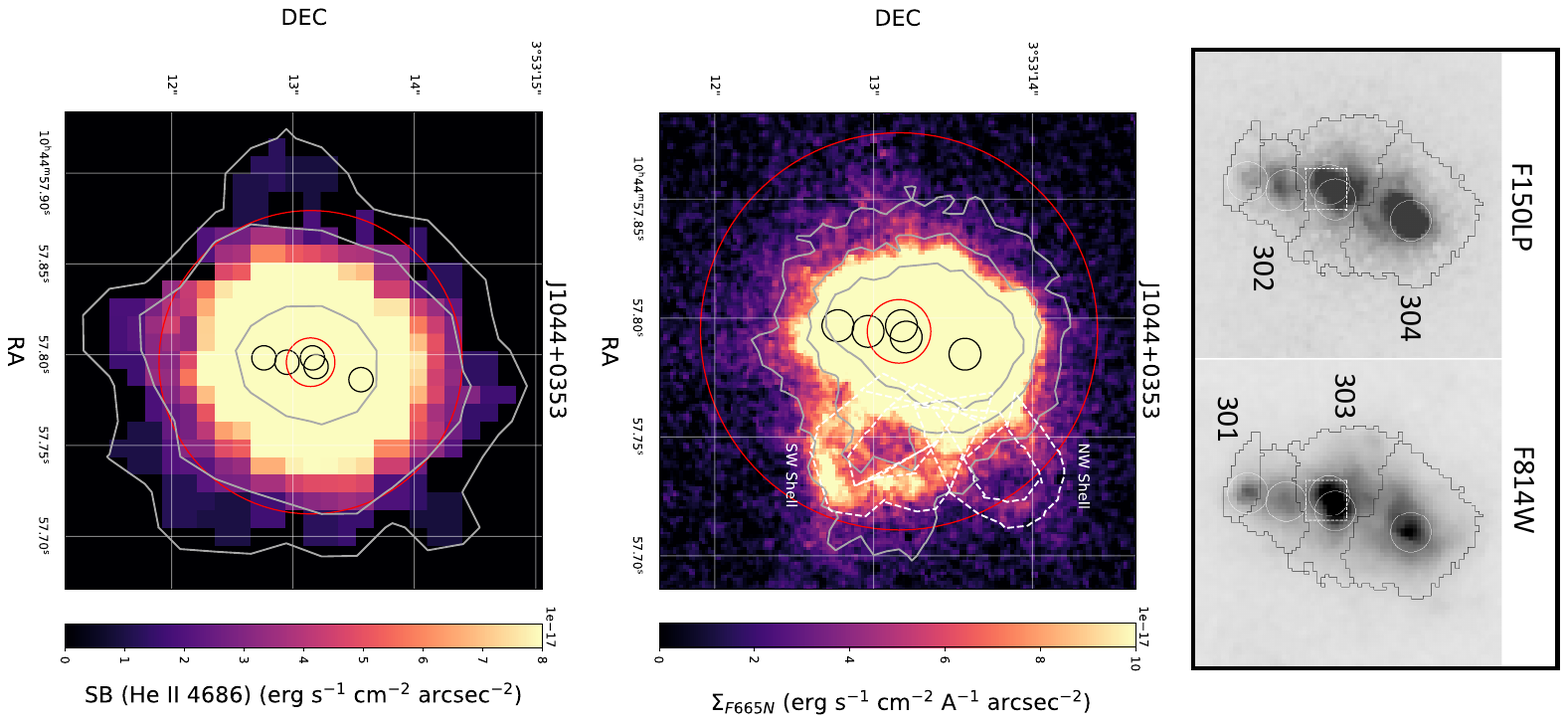}
\caption{
{\it Top:}
{\it Hubble} F150W and F814W images zoomed in on the starburst.
Small circles (0\farcs1 radius)  mark the centroids of individual star 
clusters within the four region indentified by the segmentation map.
The box marks the centroid of the \heii\ emission. 
{\it Middle:}
{\it Hubble} F665N image shows the \Ha\ emission at high resolution.
Star clusters are marked with black circles.
The northwest shell extends 395~pc (1\farcs42) from Region 303.
The brighter, southwest extended 341 pc (1\farcs23) from Region 304.
Contours compare the extended UV emission; levels are 
$6.6 \times 10^{-17} $, $1.3 \times 10^{-16}$, and
$2.7 \times 10^{-16}$ ergs~s$^{-1}$~cm$^{-2}$~\AA$^{-1}$ per square arcsecond
(uncorrected for extinction).
For scale, an 0\farcs2 radius circle is shown at the center of the 2\farcs5 red circle.
{\it Bottom:}
KCWI \heii\ $\lambda 4686$ image. The very high-ionization nebula surrounds 
the compact starburst. The outer red circle is 2\farcs5 in diameter.
Contour levels are  
$2.35 \times 10^{-18} $,
$2.35 \times 10^{-17} $, and
$2.35 \times 10^{-16} $ \flux\ per square arcsecond.
The centroid is at 10:44:57.797, +03:53:13.17 with a
positional uncertainty of one spaxel,  $\pm 0\farcs0729$ or 40~pc.
This places its origin near the two star clusters in Region 303.
The nebula is visibly elongated, and the position angle of $-12\deg$,
following the line of  young star clusters.
This morphology is consistent with multiple star clusters contributing 
$h\nu > 54.4 $~eV photons.
}
\label{fig:zoom_trio} \end{figure}

To investigate the star formation history within the starburst region, 
we detected and deblended star clusters using \textbf{photutils.segmentation}.
Different parameter combinations were explored across various filters, but they
all identified at least four subregions.  Figure~\ref{fig:zoom_trio} shows
the regions obtained using the F336W image and the following parameters:
\textbf{nlevels} = $1024$, \textbf{contrast} = 0.005, and \textbf{n\_sigma} = 3.
Table~\ref{tab:hst_phot} lists 8-band photometry for these regions. Region 303, which is the brightest 
in both F814W and F665N, consists of two clusters separated by just 24 pc.  Region 304 is slightly 
brighter than 303 in the UV filters. We fit these spectral energy distributions (SEDs) 
with stellar population synthesis models using BEAGLE \citep{Chevallard2016}. 
Our fiducial model describes each region with a single-age stellar population and adopts a
SMC extinction curve \citep{Gordon2003}.   The stellar metallicity, gas-phase metallicity,
and ionization parameter were fixed at the following values: $Z_{*} = 0.05 \zsun$, 
$Z = 0.05 \zsun$ , and $ \log(U)=-1.75$, respectively.  The fitted ages are
3.0 Myr, consistent with the \citet{Olivier2022} spectral fit.  The estimated stellar masses of Regions
303 and 304 are $3.5 \times 10^5$\msun\ and $3.2 \times 10^5$\msun, respectively. 
In order to quantify the systematic errors on these masses and ages,
we also fit the same photometry using an alternative extinction curve (\citet{Charlot_Fall2000}, CF00)
and a constant star formation rate, in total four settings.
Table~\ref{tab:hst_phot} shows that the results are not sensitive to either the reddening or the star formation
history.
We argue that the dominant systematic uncertainty is actually the initial mass function (IMF). The BEAGLE
models assume an IMF based on the Galactic disk \citep{Chabrier2003}, but the interstellar  metallicity is
much lower in J1044+0353. A higher upper mass limit would not change the estimated cluster masses significantly.

Figure~\ref{fig:zoom_trio} resolves the \Ha\ morphology within the core of the seeing-limited
PSF. Two \Ha\ shells protrude northwest of Region 303 and southwest of Region 304.  
The curved filaments defining the northwest shell converge on the Region~303 star clusters. 
The southwest shell emerges from the Region 304 cluster. 
The UV surface brightness distribution is
not symmetric about the starburst; a bright region fills the southwest shell.

We measured the far-UV flux of J1044+0353 directly from the ACS/SBC F150LP image.
The filter pivot wavelength is 1606  \AA. 
We used two apertures:  a 2\farcs2 by 2\farcs56 box around the starburst region, 
and a 8\farcs42 by 4\farcs74 box that included most of the UV light from the galaxy.
Uncorrected for extinction, the starburst and total magnitudes were m$_{150}$ = 18.70 and 18.34, 
respectively. After correcting for Galactic and internal extinction, see Table~\ref{tab:basic_properties},
the absolute magnitudes are M$_{FUV}^{SB} = -16.32 \pm 0.17$ and M$_{FUV}^{Tot} = -16.68 \pm 0.17$.
These values, listed in Table~\ref{tab:qphot}, supercede the GALEX values
which are affected by object blending.

Finally, we measured the axis ratio of J1044+0353 from the F814W isophotes. 
Despite the irregular morphology
of J1044+0353, the interstellar medium likely has a disk component;
the outflow axis is perpendicular to the major axis \citep{Peng2023}. The rotation of 
the gas disk would be just $\pm 15$ \kms, the circular velocity estimated by \citet{Xu2022}.
The violent kinematics of the bipolar outflow,  and a possible infalling gas cloud \citep{Peng2023},
mask the rotational velocity of a gas disk in the KCWI cube. 
At the F814W contour where the length of the major-axis is 2.3~kpc (8\farcs27), we find the minor-to-major
axis ratio is 0.64. If the underlying distribution of stars is disk-like, then we estimate a 
disk inclination of $50^{\circ}$ in the thin disk limit, higher for a thick disk. For example,
modeling the starlight as an oblate spheroid with intrinsic height-to-radius 
ratio of 0.5, increases the inclination to $70^{\circ}$. This edge-on orientation suggests that the
radial outflow speed is significantly higher than the line-of-sight outflow speeds, which 
are 40 - 50 \kms.

\section{Ionizing Photon Luminosities, Str\"{o}mgren Radii, and RMS Densities} \label{sec:results}

In this section, we describe the hydrogen and helium emission from the ionized gas in and around 
J1044+0353.  We describe the nebular morphology, measure the Str\"{o}mgren radii of each ionization zone,
derive the ionizing photon luminosities absorbed by the ISM at 1 and 4 Rydberg.  The luminosities and 
sizes of the \hplus\ and \heplusplus\ nebulae then determine their root-mean-square (RMS) electron density.

\subsection{Recombination Line Images} \label{sec:recombination}

Prominent emission lines from \ariv\ $\lambda \lambda 4711.35, 4740.20$ and \hei\ emission 
are visible in Figure~\ref{fig:kcwi_spec} near the 
\heii\ line. Imaging through a narrowband filter would not separate these individual lines.
In the post-starburst regions, a narrow \hei\ emission component sits in a broad, photospheric 
\hei\ absorption trough, which must be  fitted and subtracted to form a \hei\  narrowband image.
Narrow-band imaging would not measure this \hei\  line flux accurately 
because it would not recognize the depressed, local continuum level. Integral field spectroscopy 
offers the only accurate method for mapping recombination-line nebulae.

\subsubsection{\heplusplus\ Recombination}  \label{sec:he_second_component}

Nebular \heii\ $\lambda 1640$ and $\lambda 4686$ emission is produced by the recombination of 
\heplusplus, and thus demonstrates the presence of hard ionizing photons with energies 
$h\nu > 54.4$~eV.  
{\clm
The velocity width of these nebular lines is narrow, and the majority of 
local \heii - emitters, once AGN are excluded, do not present a broad emission-line component
\citep{Shirazi2012}. However, spectra of some star-forming galaxies show a broad \heii\ 
spectral feature attributed to the optically thick stellar winds of classical Wolf-Rayet stars
\citep{Crowther2007,Vacca1992}. These narrow and broad components are not
exclusive. Spatially resolved studies confirm that classical Wolf-Rayet stars power many 
of the \heii\ nebulae in M33 \citep{Kehrig2011} for example. However, the association between narrow
\heii\ emission and classical Wolf-Rayet stars does not extend to low metallicity galaxies 
\citep{Kehrig2008}, where very massive stars (VMSs), defined as those having masses 
above 100\msun \citep{Vink2011}, produce  broad, stellar \heii\ emission lines, sometimes
underneath a narrow nebular emission line \citep{Smith2016,Leitherer2018,Wofford2023}. 
}
After describing the overall morphology of the \heii\ emission in relation to the young, 
massive clusters in J1044+0353, we use our data cube to carefully examine the \heii\ line profile.

The bottom panel of Figure~\ref{fig:zoom_trio} shows the \heii\ line image. The centroid 
coincides with Region 303.  The elongation of the \heii\ contours follows the chain of 
young clusters extending from Region 301 to 304. The \heii\ emission is detected beyond
the COS aperture,  indicated by the red circle in both Figures~\ref{fig:zoom_trio} and 
\ref{fig:3color_kcwi}. A \hei\ nebula is detected in the $\lambda 4471$ line image (not shown) 
at all position angles around SB. In contrast to the \heii\ nebula, however, we also detect \hei\ 
further to the east, around FEN, where no \heii\ emission is detected. 

\begin{figure*}[ht!]
\includegraphics[angle=0,width=7in, trim = 0 380 0 0]{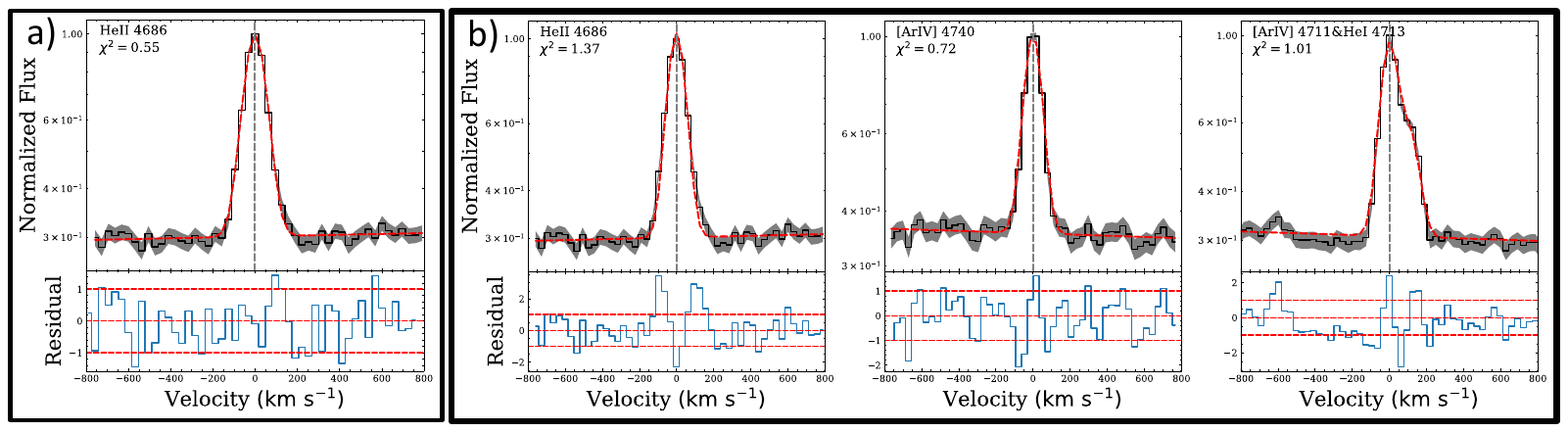}
\caption{Various Gaussian models fit to the \heii\ line profile. 
(a) The \heii\ line profile is well fit with a single Gaussian component 
($v = -31 \pm 1$ \kms,  $125 \pm 2$ \kms), but the line width is broader
than other nearby high-ionization lines. 
(b) Joint fit to to \heii, \ariv, and \hei\ $\lambda 4713.14$ lines 
with a single linewidth. The positive residual in the line wings of \heii\ demonstrate
its broader width. In the best fit to the four lines, the
\heii\ component accounts for 96\% of the total line flux
defined by the fit in Panel~{\it a}, and the error bars allow the range from from 80 to 100\%.
The broader width of \heii\ may indicate that not all of the line flux comes from \heplusplus\ recombination.
}
\label{fig:heii_fit} \end{figure*}

If recombination accounts for all the \heii\ luminosity, then the shape of the \heii\ 
profile will be related to other nebular emission lines emitted by the very-high ionization 
zone. We therefore compared the \heii\ and \ariv\ line profiles in the starburst spectrum, 
shown in the left panel of Figure~\ref{fig:heii_fit}. The \heii\ line profile is well fit by a 
single Gaussian component with standard deviation $\sigma_{tot} = 53$ \kms. After correcting 
the linewidth  for instrumental brodening, $\sigma_{inst} = 35$ \kms, the standard deviation
is $\sigma_{corr} = 40$ \kms, corresponding to 94 \kms\ FWHM.  The \ariv\ $\lambda 4740$ emission
line  profile is also well fit by a single Gauassian component, $\sigma_{corr} = 26$ \kms. 
The \ariv\ linewidth is narrower than \heii\ and  between the expectations for thermal broadening 
and turbulent broadening. 
{\clm
The larger atomic weight of argon, nine times that of helium, reduces the thermal 
linewidth of argon lines by a factor of three relative to helium lines. Assuming identical turbulent 
broadening of \heii\ and \ariv, we solved a system of two equations for the thermal velocity 
dispersion, 
\begin{eqnarray}
\sigma_{th,HeII} = \sqrt{9/8 (\sigma_{tot,HeII}^2 - \sigma_{tot,ArIV}^2)}
= 31 {\rm ~\kms}, 
\end{eqnarray}
and the turbulent velocity dispersion,
\begin{eqnarray}
\sigma_{turb} = \sqrt{(\sigma_{tot,HeII}^2 - \sigma_{inst}^2 - \sigma_{th,HeII}^2} = 25 {\rm ~\kms}.
\end{eqnarray}
This interpretation of the broader \heii\ linewidth yields a gas temperature, 
$T = 2.4 \times 10^5 {\rm ~K} ( \sigma / 31 {\rm ~km/s} )^2 (m/4 {\rm ~amu})$.
Since this implied temperature is much higher than the  measured temperature in the high ionization 
zone, $T_e \approx 2 \times 10^4$ \citep{Peng2023}, we consider an alternative explanation, namely
that the \heii\ profile includes a stellar component which does not emit \ariv.
Strong stellar winds  emit broad \heii\ lines when the helium is doubly ionized.
This \heii\ component is seen in spectra of young super star clusters in dwarf galaxies
\citep{Smith2016,Leitherer2018,Wofford2023} as well as in classical Wolf-Rayet stars. Those super
star clusters include very massive main-sequence stars rather than classical Wolf-Rayet stars 
which are evolved, helium-core burning stars \citep{Grafener2008}. 
}

Panel~{\it b} of Figure~\ref{fig:heii_fit} illustrates the larger velocity spread of 
the \heii\ emission relative to \ariv\ and \hei.  Subtraction of a joint fit to the 
\heii, \ariv, and \hei\ $\lambda 4713$ lines from the data reveals positive residuals 
extending up to $\pm 200$ \kms\ around the \heii\ line center. 
{\clm
The \heii\ linewidths of super star clusters in NGC 3125 are 1000 to 1400 \kms\ FWHM  
\citep{Wofford2023}, much larger than our proposed second component in J1044+0353. 
One reason for the difference could be the lower gas-phase metallicity of J1044+0353,
which is just one-seventh the LMC-like metallicity of NGC 3125.
As stellar metallicity decreases, stellar rotation is expected to increase, thereby 
decreasing the effective escape speed and increasing the effective Eddington ratio  
\citep{Meynet2002}.  At stellar metallicities of $0.1 \zsun$, i.e. slightly higher
than the gas-phase metallicity of J1044+0353, theoretical models predict \heii\  
line widths  of just 490 \kms\ FWHM for WNh stars, a type of VMS \citep{Grafener2015}.
Considering the very young age of the J1044+0353 starburst, we conclude that VMS are the most likely 
source of the second, broader component that we identify in the \heii\ line profile.
}

{\clm
\citet{Grafener2015} argue that heii\ emission from low-metallicity stellar winds could 
be mistaken for nebular emission due to the low wind velocities, but this is not 
the case in J1044+0353. The \heii\ luminosity from J1044+0353 is dominated by nebular
emission. This  very-high ionization nebula subtends a larger solid angle than the string
of compact clusters in Figure~\ref{fig:zoom_trio}, so the \heii\ emission cannot be 
dominated by stellar emission.
The detection of collisionally excited \ariv\ lines, which have no stellar counterpart, 
further demonstrate the presence of a very-high ionization nebula. The amplitudes of the \heii\ and 
\ariv\  $\lambda 4711$ line are equal across most of the nebula; the only exception is a spectrum
extracted from a seeing limited aperture centered on Region~303, where the \heii\ line has a 
larger amplitude than the \ariv\ $\lambda 4711$ line. 
}

Photoionization by an AGN, essentially a point source, is not consistent with the overall elongation of 
the nebula in Figure~\ref{fig:zoom_trio} or the optical emission-line ratios. The high production 
efficiency of \heplus -ionizing photons appears to be a property of very young, low-metallicity stars, 
a population which likely includes VMS in J1044+0353. The presence of this stellar \heii\ component lowers
 the  inferred photoionization rate, \qheplus, slightly.   We estimate this correction by noting that
the residual flux in Panel~{\it b} accounts for $4^{+16}_{-4}\%  $ of the  total line 
flux in Panel~{\it a}. Hence $\qheplus_{corr} = 0.96^{+0.04}_{-0.16} \qheplus$, where
\qheplus\ is our measured value below.

\subsubsection{Hydrogen Recombination}

We present a global view of the hydrogen recombination-line nebula in 
Figure~\ref{fig:3color_kcwi}. The faint shells contribute very little of the total \Hb\ luminosity.  
Most of the recombinations occur near the compact starburst.  There, the \Hb\ isophotes are round 
out to 1.2~kpc (4\farcs3). It is therefore reasonable to describe the nebula photoionized by the
starburst as a galactic-scale  Str\"{o}mgren {\it sphere}. 

On larger scales, the \Hb\ surface brightness profile depends on position angle. 
In projection on the sky, the starburst to shell separation reaches $\rx = 3.3$~kpc (12\farcs0). 
The faint shell protruding directly north of SB extends to $R = 1.99$~kpc (7\farcs15). To the west 
of SB, low surface brightness emission extends to at least 2.5~kpc. 
In the east, the \Hb\ contours and nearly circular around FEN. The presence of 
stars hot enough to ionize hydrogen indicates ongoing star formation in FEN. The Balmer 
absorption in the FEN line profiles, however, indicate the ratio of Type A stars to Type O
stars is larger than its ratio in SB. \cite{Peng2023}  modeled the star formation history
as a burst 20 Myr ago plus a constant component.
In contrast to the Str\"{o}mgren  spheres powered by FEN and SB, the peanut shape of
the \Hb\ contours draws attention to the minimum in the \Hb\ emission between them. 
The faintest \Hb\ contours follow the extended \oiii\ filaments visible in Figure~\ref{fig:3color_kcwi}.
These loops are shells of gas which the outflow has swept out of its path.

\subsection{Photometry}

The compact SB is tiny compared to the nebula it ionizes, a geometry similar to the
\hii\ regions around massive stars, but on a global scale. 
The centroid of the \heii\ emission lies in Region~303, and we measured integrated emission-line 
fluxes in a series of concentric circular apertures around the centroid of the \heii\ emission
In each aperture, we measured the flux in the
recombination-line images, and we extracted a spectrum from which we directly measured 
the integrated line profile.  These methods gave consistent fluxes. 
The photometric growth curve for the \heii\ $\lambda 4686$ emission reaches a well-defined 
asymptotic level, so the total flux is well defined.

\begin{figure}[h]
\includegraphics[angle=0,width=3.5in]{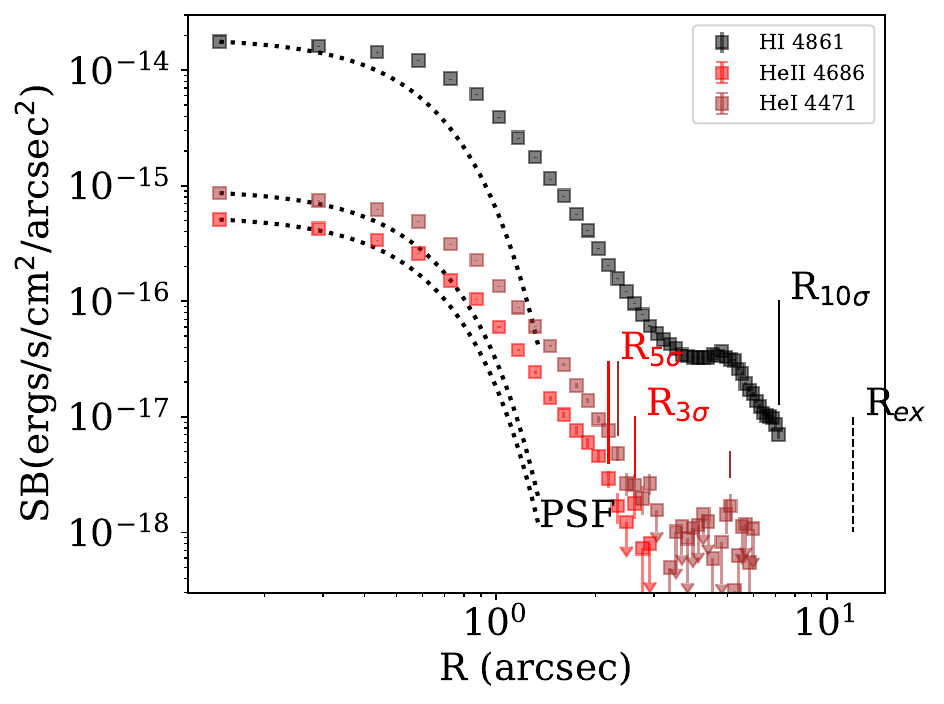}
\caption{Surface brightness profiles.
The post-starburst region is detected at radius 5\farcs1 in the \hei\
$\lambda 4471$. The \Hb\ emission is detected at $10\sigma$ at
radius 7\farcs14, the largst circular aperture with complete coverage.
Along the long axis of the mosaic, the \Hb\ emission is detected to
radius 12\farcs0.
}
\label{fig:sb_profiles} \end{figure}

Figure~\ref{fig:sb_profiles} presents the radial surface brightness profiles. 
The ionization structure follows the expectation for a centrally concentrated source; 
the very-high ionization zone is smaller than the other zones in Figure~\ref{fig:sb_profiles}.
The seeing-limited PSF shapes the profile at radii less than 1\farcs0. 
The surface brightness profile falls steeply between 1\farcs0 and 2\farcs0.
The \hei\ surface brightness declines at a rate similar to the \heii\ 
emission out to 2\farcs0, beyond which the nebula centered on the FEN region flattens the
profile.   The edge of the \heplusplus\ Str\"{o}mgren sphere is  therefore sharper than the 
boundary of the \heplus\  nebulae. 
The inner \Hb\ profile can be described by the same exponential scale length
as \heii\ and \hei, but contribution from the FEN region flatten the profile at larger radii.
The aperture photometry describes the nebular structure out to radii where the isophotes can
be described as round but does not capture the filamentary structures at larger radii.

At many position angles, the \Hb\ emission extends to the edge of the mosaic cube. The
profile in Figure~\ref{fig:sb_profiles} is truncated at the largest circular aperture 
that fits within the rectangular mosaic. The slope of the \Hb\ growth curve is very close
to zero at these distances, so the filamentary emission would not increase the total observed 
\Hb\ flux significantly. In Table~\ref{tab:qphot}, we list two 
radii determined from visual examination of the binned recombination-line images. The smaller 
one, $R_{c}(X)$ where $X$ identifies the  emission line, describes the radius where the shape 
of the intensity contours transitions from largely circular,  to mainly filamentary, structures.
The scale of the more filamentary emission is position angle dependent and limited by the 
mosaic field-of-view for \Hb.  We interactively examined the spectrum across the nebula to
determine the maximum extent of the \Hb\ emission. We denote the maximum projected radius relative to 
the starburst  as $R_{x}(X)$.  These radii are marked in Figure~\ref{fig:sb_profiles}. 
These visual classifications correspond to S/N ratios of 5 and 3 roughly.
We adopt the  $R_c$ values as the best  direct measure the Str\"{o}mgren radii.  
The \heii\ emission is detected within $\rc = 610$~pc (2\farcs2) of the compact
starburst at all position angles and extends to $\rx = 730$~pc to the northwest and 
the southeast.

\subsection{Ionizing Photon Luminosities}  \label{sec:luminosities}

In this section, we assume the nebula is photoionized. Then the recombination rates 
determine the ionization rate of H and \heplus\ in the galaxy. 
We calculate \qh\ and \qheplus, the photon luminosities at $h\nu \ge\ 13.6$ eV and $h\nu \ge\ 54.4$ eV
respectively, from the recombination-line luminosities measured from the KCWI data cubes. These
integrated luminosities are listed in Table~\ref{tab:qphot} and are larger than the values 
measured from fixed apertures such as the SDSS fiber spectrum. 
We discuss modifications for the possible stellar contribution to the \heii\ luminosity
in Section~\ref{sec:discussion}.

Following recombination of an electron with a hydrogen or helium ion, the electron 
cascades downward to lower energy levels producing the recombination line emission.  
In photoionization equilibrium, the \hi\ and \heii\ recombination-line luminosities
determine the photoionization rates of H atoms and \heplus\ ions, respectively.
The luminosities of \Hb\ and\ \heii\ $\lambda 4686$ therefore measure the
photon luminosity absorbed by the ISM at energies exceeding 13.6 eV and 4 Rydberg 
(54.4 eV), respectively. Correction for the leakage of ionizing radiation from the galaxy
then yields the intrinsic EUV continuum level at these energies, \qh\ and \qheplus\ respectively.

In photo-ionization equilibrium, the hydrogen ionizing photon luminosity 
follows directly from the luminosity of a recombination line, 
\begin{eqnarray}
\qh = \frac{\alpha_B(H,T)}{\alpha^{eff}_{H\beta} h \nu_{H\beta}} \frac{L_{H\beta}}{(1-\fesc)},
\label{eqn:qh} \end{eqnarray}
where $(1 - \fesc)$ denotes the fraction  of the ionizing photon luminosity absorbed by the ISM.
The on-the-spot approximation for the diffuse radiation field \citep{Osterbrock2006} may break down
along some channels through the ISM, but the recombination emission will come from gas that presents
a significant cross section to the ionizing radiation.
We therefore adopt Case~B recombination coefficients \citep{Pequignot1991,Storey1995} at 
$T_e = 2.0 \times 10^4$~K, the electron temperature measured  in the highly-ionized O$^{+2}$ 
zone \citep{Peng2023}.
Inserting the recombinatino rates in  Eqn.~\ref{eqn:qh}, the ionizing photon 
luminosity becomes
\begin{eqnarray}
\qh = 4.35 \times 10^{52} {\rm ~s}^{-1} \frac{L_{H\beta}}{2 \times 10^{40} {\rm ~ergs~s}^{-1} (1 - \fesc)},
\label{eqn:qh_units}.
\end{eqnarray}
which evaluates to  $\qh = 5.04 \pm 0.48 \times 10^{52} (1-\fesc)^{-1}{\rm ~ s}^{-1}$ for the
extinction-corrected luminosity of J1044+0353.

The \heplus-ionizing continuum at energies $\ge\ 54.4$~eV can be directly measured from the
luminosity  of the \heii\ $\lambda 4686$ recombination line. In photoionization equilibrium, we have
\begin{eqnarray}
\qheplus = \frac{\alpha_B(\heplus,T)}{\alpha^{eff}_{\lambda 4686} h \nu_{4686}} \frac{L_{\lambda4686}}{(1-\fescHe)} \\
                       \nonumber
         = 4.92 \times 10^{50} {\rm s}^{-1} \frac{L_{\lambda4686}}{4 \times 10^{38} (1 -\fescHe)},
\label{eqn:qhe_units} \end{eqnarray}
giving an ionizing photon luminosity $\qheplus = 4.89 \pm 0.47 \times 10^{50}(1-\fescHe)^{-1}$~s$^{-1}$
for J1044+0353. 
The result is sensitive to the electron temperature, and we
note that reducing the electron temperature from $2.0 \times 10^4$~K to $T_e = 1.0 \times 10^4$~K yields
the coefficient in Equation~\ref{eqn:qhe_units} 1.23 times higher.
We introduce \fescHe\ to distinguish the escape fraction at $h\nu \ge 4$ Rydberg from \fesc\ at the Lyman limit.

The spectral bandpass also includes several \hei\ recombination lines,
the strongest of which is \hei\ $\lambda 4471$. The size of the \heplus\ zone is 
sensitive to the temperature of the stars because hydrogen and helium present similar
cross sections at the helium ionization edge of 24.6~eV \citep{Osterbrock2006}.  
The more abundant H atoms make up for the larger cross-section of \heplus\ per ion.
It follows that the photons produced by recombinations to He$^0$, however, do not 
necessarily photoionize another helium atom. 
No simple relation exists to convert \hei\ line luminosities
to the ionizing photon luminosity at 24.6~eV \citep{Schaerer1998}.

\subsection{Root-mean-square Electron Density} \label{sec:ne_rms}

The overall size of the \Hb\ and \heii\ nebulae provides important constraints on the 
structure of the ISM. The ionization front driven by the young star clusters 
stalls when the ionized volume becomes large enough that the recombination rate balances
the ionizing photon luminosity. Even when the emitting clouds have high  electron density $n_e$,
nebulae can grow to large sizes because the clouds typically fill a tiny fraction of the geometrical
volume.

The ionized volume is the geometrical volume multiplied by the volume filling factor, \fv, 
of the emitting clouds \citep{Stromgren1948}.
For a spherical nebula that volume is $\frac{4 \pi}{3} \rhplus^3 \fv$, 
and we write the Str\"{o}mgren radius as
\begin{eqnarray}
\rhplus = \bigg( \frac{3 \qh (1-\fesc)}{4 \pi \alpha_B({\rm H^0},T) n_e n_p}  \bigg)^{1/3} \fv^{-1/3},
\label {eqn:rh} \end{eqnarray}
in order to show the inverse dependence on the filling factor.  
We assume \fv\ is constant throughout the nebula, then the quantity $\fv n_e^2$
defines the mean-square electron density, 
\begin{equation}
\nmsq \equiv \fv n_e^2.
\label{eqn:fv} \end{equation}

Using Eqn.~\ref{eqn:qh_units}, we can express \qh\ directly in terms of the observed \Hb\
luminosity, L(\Hb). 
Solving Equation~\ref{eqn:rh} for the root-mean-square (RMS) density, we then
obtain the RMS density in terms of the measured size and luminosity
of the nebula,
\begin{eqnarray}
\nrmsq = 1.22 {\rm ~cm}^{-3} \bigg( \frac{1 {\rm ~kpc}}{\rhplus} \bigg)^{3/2}  
                       \bigg( \frac{L(\Hb)}{10^{40} {\rm ~ergs~s}^{-1}}  \bigg)^{1/2} \times ~~~ \\
                       \nonumber
   \bigg( \frac{1.610 \times 10^{-14} {\rm ~ergs~cm^3~s^{-1}} }{\alpha_{H\beta}^{eff}({\rm H^0},T)} \bigg)^{1/2} 
   \bigg( \frac{n_e/n_p}{12/10} \bigg)^{1/2}.
\label{eqn:rms} \end{eqnarray}
We adopted an electron temperature $T_e = 2.0 \times 10^4$~K, appropriate
to the O$^{+2}$ zone in J1044+0353; cooler temperatures of $1.5 \times 10^4$~K
(or $1.0 \times 10^4$~K) would lower the coefficient to 1.07 (or 0.89, respectively).

The estimated RMS density in the galaxy depends on the aperture. For the H$^+$ Str\"{o}mgren sphere, 
Table~\ref{tab:qphot} lists $L(\Hb) = 2.54 \times 10^{40}$ ergs~s$^{-1}$ and \rc(\hplus) = 1990~pc.
Substitution of these values  into Eqn.~\ref{eqn:rms} gives a global RMS density of  0.69~cm$^{-3}$. 
The J1044+0353 \Hb\ emission is strongly concentrated near SB, and roughly 82\% of that \Hb\ luminosity 
is emitted within the much smaller aperture describing the He$^{+2}$ Str\"{o}mgren sphere,
$\rc(\heplusplus) = 610$~pc. Eqn.~\ref{eqn:rms} therefore indicates a higher RMS density,
$\nrmsq =  3.7$~cm$^{-3}$, for the \Hb\ emission from the very-high ionization zone.

In close analogy to Eqn.~\ref{eqn:rh}, the photon luminosity at energies $ \ge\ 54.4$~eV
determines the volume of very-highly ionized clouds in photoionization equilibrium. 
\begin{eqnarray}
\frac{4 \pi}{3} \rheplusplus^3 \fvh = \frac{\qheplus (1 - \fesc54)}{ \alpha_B({\rm He^+},T) n_e n_{He+2}}.
\label {eqn:rhe} \end{eqnarray}
We adopt a fiducial ratio of H-to-He atoms of 10:1 by number, consistent with the 12 
electrons per 10 protons adopted in Eqn.~\ref{eqn:rms}. This helium abundance is 
equivalent to a helium mass fraction $Y = 0.286$, a bit higher than the lower
bound set by the primordial fraction,  $Y = 0.2470$ \citep{Planck2020}.
We assume that helium is fully ionized in the very-high ionization zone.
A photon luminosity \qheplus\ then ionizes a volume of radius
\begin{eqnarray}
\rheplusplus= 208 {\rm ~pc} \bigg( \frac{\qheplus}{10^{50} {\rm ~s}^{-1}} \bigg)^{1/3} 
                       \bigg( \frac{1 {\rm ~cm}^{-3}}{\nrmsq}      \bigg)^{2/3} \times\ ~~~ \\
                       \nonumber
                       \bigg( \frac{1/10}{n(He^{+2})/n_e}       \bigg)^{1/3} 
                       \bigg( \frac{9.03 \times 10^{-13} {\rm ~ergs~cm}^3~{\rm s}^{-1}}{\alpha_{B}(He^+,T)} \bigg)^{1/3}
                       \fvh^{-1/3},
\end{eqnarray}
where we have evaluated the recombination coefficient at a fiducial temperature
of $T_e = 2.0 \times 10^4$~K.\footnote{
          The radius would be 93\% (83\%) of this value  at temperatures
          of $1.5 \times 10^4$~K or $1 \times 10^4$~K, respectively.}
The filling factor of the very highly ionized gas, \fvh, is then defined by the ratio
of the RMS density derived from the \heii\ emission and the local, physical density of the clouds.

Solving for the RMS density in the \heplusplus\ zone and substituting for \qheplus\ using
Eqn.~\ref{eqn:qhe_units}, we find
\begin{eqnarray}
\nrmsq = 0.25 {\rm ~cm}^{-3} \bigg( \frac{600 {\rm ~pc}}{\rheplusplus} \bigg)^{3/2} 
\bigg( \frac{L_{\lambda4686}}{10^{38} {\rm ~ergs~s}^{-1}}  \bigg)^{1/2} \times\ ~~~~ \\
\nonumber
\bigg( \frac{n(H)/n(He)}{10}  \bigg)^{1/2}
\bigg( \frac{10/12}{n_p/n_e} \bigg)^{1/2}.
\label{eqn:rms_he2} \end{eqnarray}

For a \heii\ luminosity $4 \times 10^{38}$~ergs~s$^{-1}$ and \rc(\heplusplus) = 610~pc,
this expression yields an RMS density \nrmsq = 0.50~cm$^{-3}$ 
for the very-high ionization zone.

In summary, our analysis of the integrated spectrum extracted from the fixed aperture defined
by \rc(\heplusplus) finds different RMS densities for the \heii- and \Hb-emitting regions.
This result has implications for the physical structure of the \hii\ region. 
In the next section, we measure the local electron density in the different ionization zones.
Then, in Sec.~\ref{sec:fill_factor},  we compare these physical densities and the RMS 
densities estimated here, thereby measuring the filling factor in each ionization
zone via Eqn.~\ref{eqn:fv}.

\section{Global Ionization and Density Structure} \label{sec:structure}

The mosaic cube detects  \oiii\ $\lambda \lambda 4959, 5007$ emission 
across the entire field of view shown in Figure~\ref{fig:3color_kcwi}.
The emission-line nebula therefore extends further beyond the starlight
than a casual comparison to Figure~\ref{fig:kcwi_fov_duo} suggests. 
Large spatial gradients in the intensity ratio of the  \oiii\ doublet relative to the 
\oii\ $\lambda \lambda 3726, 29$ doublet have two practical implications: 
(1) the area over which the \oii\ doublet can be used to measure the electron density 
is restricted;  and (2) we can map out the pathways of Lyman continuum leakage.
In this Section, we use the collisionally excited lines to describe the physical 
properties of the ionized gas:  
electron density and volume filling factor in Section~\ref{sec:ne}, and the
ionization parameter and LyC leakage in Section~\ref{sec:u}.

\subsection{Density Structure} \label{sec:ne}

Densities derived from ions with very different ionization potentials provide insight into 
the structure of the ISM. In this section, we derive the gas density in low and high 
ionization zones using  density-sensitive ratios of emission lines.  We then argue
that the volume filling factor of these clouds is quite low, a conclusion that follows from the 
observed sizes of the spatially resolved Str\"{o}mgren radii.

\subsubsection{Density Gradients with Ionization Zone}

The intensity ratio of lines collisionally excited from the same ground state depends only on the 
relative statistical weights of their upper levels when the gas density is well below the critical 
densities of the transitions. In the density range where collisional de-excitation rates are important 
in one of the two lines, the line ratio directly measures the electron density. 
The bandpass covered by 
our KCWI observations covers several density-sensitive emission-line doublets. The \oii\ doublet, 
hereafter ${\rm O2DR~} \equiv F(\lambda 3729) / F(\lambda 3726)$, probes the low-to-intermediate ionization 
zones where ions with ionization potentials between 13.6 - 35.1 eV coexist. We applied 
PyNeb \citep{Luridiana2015}  to derive densities from line ratio measurements..

In order to compare the density in different ionization states,
we will discuss a spectrum extracted from a circular aperture of radius \rc(\heplusplus).
To place these densities in the broader context of the galaxy, we note that
the SB region has higher $n_e$ than FEN, NEN, or MEN.  In addition, 
spectra extracted from larger apertures have a higher O2DR, equivalent to a lower $n_e$.
Within \rc(\heplusplus) aperture, the \oii\ density is $185^{+9}_{-11}$~cm$^{-3}$. 
In the larger aperture defined by \rc(\hplus), but still centered on SB,
the density is $161^{+7}_{-10}$~cm$^{-3}$. 

Figure~\ref{fig:kcwi_spec} shows the strong nebular [Ar IV] $\lambda  \lambda 4711, 40$
emission.  The \ariv\ doublet ratio $F(\lambda 4711) / F(\lambda 4740)$ is sensitive to the electron 
density in regions of the ISM where ions with ionization potentials between 40.7 - 59.8 eV are found;
a region including both the high-ionization zone where \oiii\ is emitted (35.1 - 54.9 eV) and the 
very-high ionization zone emitting \heii. The J1044+0353 KCWI cube detects the \ariv\ doublet 
at high S/N ratio across the entire \heii -  emitting zone, as defined by \rc(\heplusplus) in 
Table~\ref{tab:qphot}.

We measured the \ariv\ line
strengths by fitting Gaussian profiles. Table~\ref{tab:qphot} lists the results.  Applying PyNEB to these
line ratios, we computed an \ariv\ density of $1440^{+44}_{-128}$~cm$^{-3}$. The \oii\ density 
was lower,  $185^{+9}_{-11}$~cm$^{-3}$, in the same spectrum. This result demonstrates that the density in 
the very-high ionization zone is larger than the density in the low-to-intermediate ionization zone; 
the large difference provides insight into the structure of the ISM.
We do not detect  significant spatial variations in $n_e(\ariv)$ within \rc(\heplusplus).

Intermediate ionization zone (24.4 - 47.9 eV) 
densities measured via the [C III] $\lambda 1907$ / C III] $\lambda 1909$ ratio typically follow a similar trend
where the \ciii\ densities are higher than \oii\ densities \citep{James2014}. 
The KCWI spectra also cover the \cliii\  $\lambda \lambda 5517, 37$ doublet, a density diagnostic probing
the intermediate-to-high ionization zone (23.8 - 39.6 eV). In our KCWI spectra, the \cliii\ doublet lands 
in a spectral region attenuated by the dichroic beam splitter. As a result, the S/N ratio is much lower 
than we obtained for  the other density measurements, and we do not discuss it further here. 
In the next section, we provide further insight into the structure of the low-ionization 
and very-high ionization gas clouds by computing their volume filling factors.

For completeness, we  note that recombination of oxygen ions can lead to \oii\ 
emission from very dense regions \citep{Osterbrock2006}, but the gas density in J1044+0353 is not high enough 
for recombination to enhance \oii\ emission.

\subsubsection{Density Gradients with Distance from SB}

The density gradient across J1044+0353 may be influenced by several factors
including gas flows and hydrostatic equilibrium.  To gain further insight about
the structure of the ISM, we mapped the electron density across this low mass galaxy.
Toward the northwest, west, and southwest of the starburst, the density declines with radius. 
The density gradient  is shallowest to the east (PA $\approx 90\deg$), where it remains constant 
over more than a kiloparsec, and steepest to the west (PA $\approx   270\deg$).
The galactic center of mass is not well constrained, but the high densities suggest
it lies somewhere in the region between SB and MEN.

\begin{figure}[h]
\includegraphics[angle=0,width=3.5in]{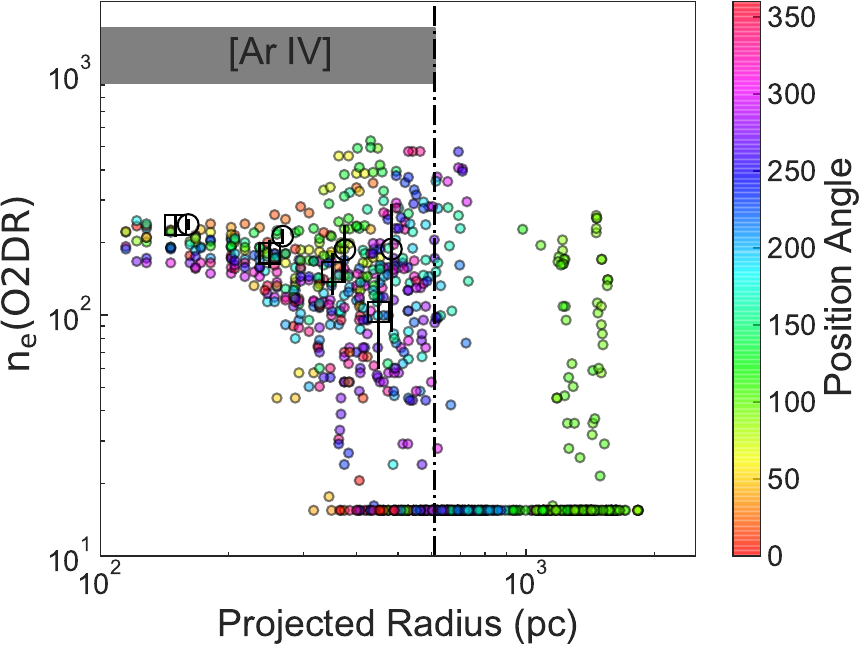}
\caption{Radial density profile. Projected distance was computed relative to the 
compact starburst.  Electron density calculated from \oii\ $\lambda  \lambda 3726, 29$ 
doublet ratio at $T_e = 1.5 \times 10^4$~K. The gray band shows the much higher density
derived from \ariv\ in the very-high ionization zone.
Median values are computed separately for PA = 10$^{\circ}$ to -170$^{\circ}$ counter-clockwise 
(large black squares), and position angles from 10$^{\circ}$ to 190$^{\circ}$ clockwise 
(large black circles).
The standard deviation (black error bars) illustrate the large spread in density.
The density is roughly constant in the direction of the post-starburst region.
In contrast, the density gradient is more negative in the hemisphere toward the west.
The point-spread function flattens the density 
profile $n_e(r)$ within 250~pc ($2 \times$~FWHM) of SB. 
}
\label{fig:ne_map} \end{figure}

We measured the O2DR for a set of spectra extracted from apertures defined by an
adaptively binned map of the \oii\ emission.  Figure~\ref{fig:ne_map} plots the electron 
density derived from these O2DR measurements as a function of the projected distance 
from SB. The dispersion in the $n_e$ values increases with separation from SB. We have
labeled the points by position angle in order to contrast the density gradients in
different directions.

\subsubsection{Volume Filling Factor}  \label{sec:fill_factor}

The 3D spectroscopy made it possible to resolve the spatial extent of the ionized nebula
and its inner region where helium is doubly ionized. These measurements determine the
RMS electron densities over the associated macroscopic volume. The RMS densities
are much lower than the local density in the clouds because the emitting clouds do not fill 
the Str\"{o}mgren sphere. We used density-sensitive line ratios to measure the physical density 
in the ionized gas clouds. Comparison to the RMS density then defines their volume filling factor, 
$\fv \equiv \frac{\nmsq}{ n_e^2}$.
Table~\ref{tab:basic_properties} summarizes the density measurements,
and inspection shows all the physical densities are higher than any of the RMS densities.

Volume filling factors for \hii\ regions are typically derived from \oii\ or \sii\
densities because high-ionization lines are often weak and/or in the rest-UV spectrum.
Over the full Str\"{o}mgren sphere, \rc(\hplus),  our measurements yield 
\fv  $ \approx  (0.69 / 161)^{2} \approx 1.8 \times 10^{-5}$. 
These filling factors are very low compared to Galactic \hii\ regions, 
where the filling factors are a few percent \citep{Osterbrock2006}. 
Using the KCWI cube, spectral extraction from smaller apertures leads to higher
filling factors because the gradient in the physical density is shallower than the 
surface brightness profile of the recombination lines. Within the \heii\ Str\"{o}mgren
sphere of J1044+0353, for example, we estimate \fv  $ \approx  (3.7 / 185)^{2} \approx 4 \times 10^{-4}$.

The strong \heii\ and \ariv\ emission of extreme emission-line galaxies (EELGs) 
make it possible to measure another  filling
factor, \fvh. For this very-high ionization gas, we find $\fvh \approx (0.5 / 1440)^2 \approx 
1.2 \times 10^{-7}$. Within the \rc(\heplusplus) aperture, we then have $\fvh / \fv \sim
3 \times 10^{-4}$.  The very-high ionization emission comes from 3\% the photoionized gas
by volume.

\subsection{Ionization Structure}

In the canonical picture of a spherical \hii\ region, the ionization structure is layered in concentric 
shells, the higher ionization zones lying interior to the lower ionization zones. When this model is applied 
to the global structure of a galaxy, emission from the low ionization-state gas comes from a zone located
at larger radii than the very-high and high ionization zones. Photoionization models with four zones 
can fit most of the emission lines (not \heii) in the J1044+0353 spectrum \citep{Berg2021}. Here, however, 
we argue that this onion-like structure does not provide an adequate description of the global 
ionization structure in J1044+0353. Specifically, the \oii\ emission does not come from a shell. 
If it did, then we would see the same shell at all projected radii, but we find a density gradient.
The emission from very-high ionization states is centrally concentrated, but, in contrast to the 
canonical picture, we find lower density, less ionized gas co-exists with the highly-ionized, 
high-density gas. 
Neither of these components of ionized gas fill the majority of the volume, so
the observed ionization structure requires a multiphase medium. Consistent with this picture, we note that
radiation pressure dominated \hii\ regions have a wide range of gas
densities (and therefore ionization parameters) within individual clouds \citep{Stern2016}.

A population of clouds surrounded by much hotter (and lower density) gas offers a more realistic 
schematic. The dynamical importance of both the hot phase and radiation pressure have been discussed 
in the context of quasar outflows \citep{Stern2016}. Propagation of an ionization front through the 
clouds stratifies the ionization structure on the spatial scale of individual clouds. The ionization 
state decreases with increasing depth into a cloud. With this picture in mind, the
\heii\ emission comes from the cloud surfaces facing the compact SB, and this very-high ionization 
zone is ionization-bounded within a typical cloud. The \oii\ emitting volume of a typical cloud
is higher than the \heii\- emitting volume, even within \rc(\heplusplus).  This structure
requires very dense gas clouds in addition to a hard ionizing spectrum.  We discuss the \cite{Stern2016}
models in the context of J1044+0353 further in Sec.~\ref{sec:discussion}.  In this section we use traditional 
photoionization models to interpret emission-line ratios. We select plane-parallel, rather than spherical
geometry, and picture the resulting ionization gradients as describing individual clouds.  Varying
the ionization parameter of the cloud then describes how radial gradients in cloud density and 
radiation intensity change emission-line ratios.

\subsubsection{Ionization Parameter} \label{sec:u}

The value of the ionization parameter, the number of ionizing photons per hydrogen atom, causes 
significant variations in emission-line ratios among \hii\ regions having the same gas-phase 
metallicity and ionizing spectrum. The strongest feature in rest-optical spectra is the strength of
collisionally-excited lines from O$^{+2}$ relative to O$^{+}$. At metallicities below solar,
the \oiii\ / \oii\ flux ratio, or O32, increases linearly with the ionization parameter \citep{Kewley2002}. 
Previous work identified J1044+0353 as a candidate Lyman continuum leaker based on the high O32 ratio of 
the starburst spectrum \citep{Berg2022}. Mapping the projected ionization structure provides new insight 
about the ionization structure of the nebula and the anisotropic escape of LyC radiation.
In this paper, we define the O32 ratio by the strength of the 
\oiii\ $\lambda \lambda 4959, 5007$ doublet  relative to the \oii\ $\lambda \lambda 3726, 3729$ doublet.  
For comparison to papers that exclude \oiii\ $\lambda 4959$ from the numerator of O32, we multiply their 
ratios by four-thirds (an addition of 0.125 dex) before comparing to our O32 measurements.

\begin{figure}[h]
\includegraphics[angle=0,width=3.75in]{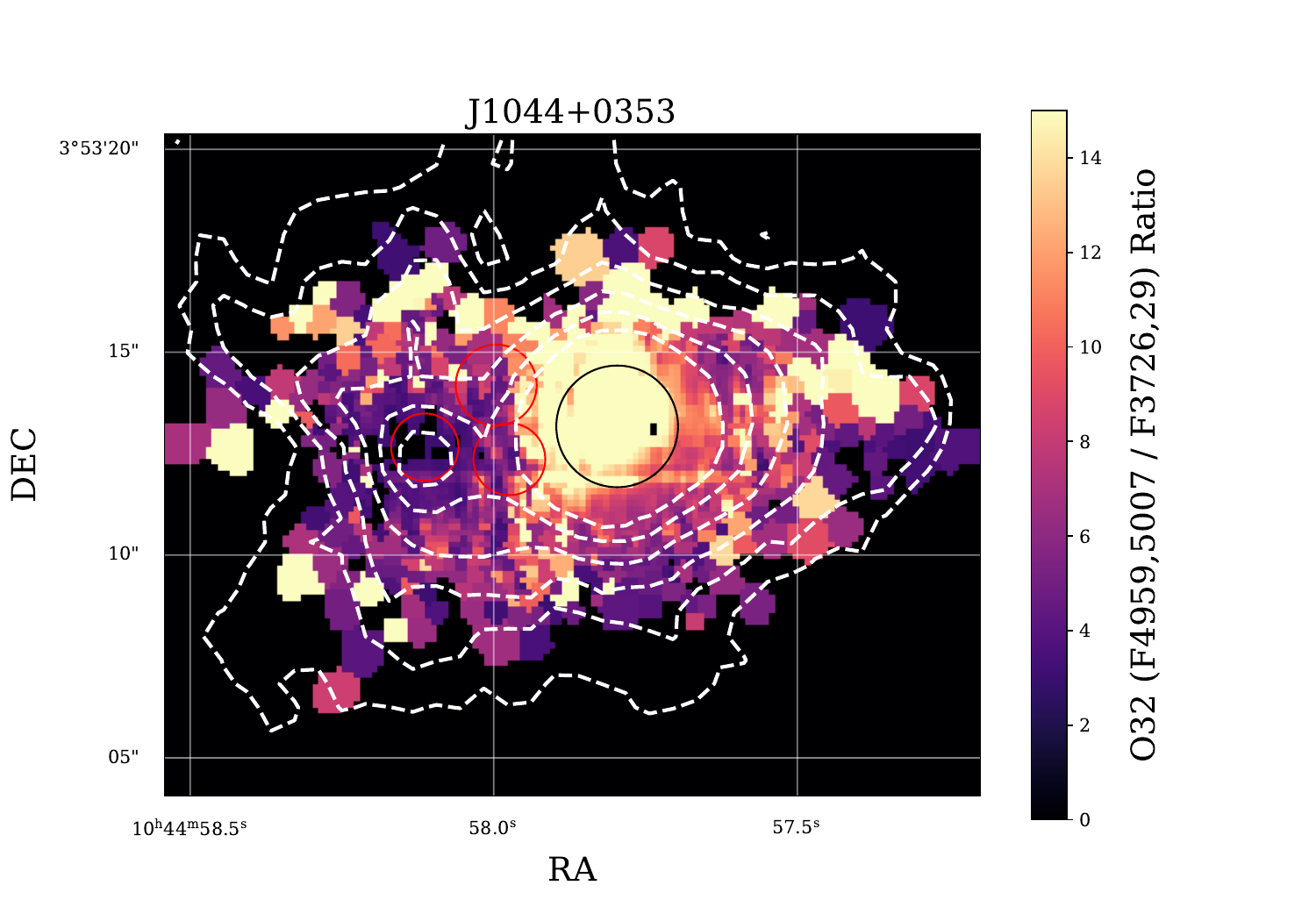}
\caption{The O32 line ratio map. The \oiii\ doublet is detected all the way to 
the edge of the map; {\clm but the map is black where no \oii\ emission is detected. }
The red and black circles mark the location of the starbust and FEN, respectively.
The negative radial gradient to the east and west of the starburst are typical of 
ionization-bounded galaxies. In contrast, north-northeast of the starburst, the 
O32 ratio remains high to the edge of the frame, a sign of a density bounded nebula.
{\clm
The O32 ratio exceeds 18  in the region between the two outermost contours 
to the north-northeast, and we call this region the ionization cone.}
The contours denote the \oiii\ $\lambda 5007$ surface brightness (by factors of two 
from $1.5 \times 10^{-17}$ to $4.8 \times 10^{-16}$ \flux\ per square arcsecond.
See Section~\ref{sec:u} for further details.
}
\label{fig:O32_j1044} \end{figure}

Figure~\ref{fig:O32_j1044} presents our map of the O32 ratio. The mean O32 ratio within 0\farcs9 of SB is 18.
In the post-starburst region around FEN, the O32 ratio is much lower, roughly 3.
The O32 values are highly position-angle dependent  in the  annulus between 0\farcs9 and 1\farcs5 surrounding SB.
The minimum ratio is O32 $= 10$ at PA=225$^{\circ}$. The high O32 ratio to the north north-east of SB 
(PA = 44 to -11) defines an ionization cone. The O32 values  between position angles of  0 and $25^{\circ}$  
remain between 13 and 18 out to projected radii of at least  4\farcs2 (1.2 kpc). The solid angle subtended by 
this cone  intersects the bipolar outflow identified by \cite{Peng2023}. Its size is roughly 20\% of $4\pi$ 
steradian near SB.
At larger radii, the O32 measurements are truncated 
because no \oii\ emission is detected.  Using the \oiii\ contours to extend the
ionization cone, we measure a lower limit, ${\rm O32} > 18 $, at the northern edge of 
the mosaic. This definition of the ionization cone is extremely conservative.
At nearly every position angle, we detect the \oiii\ doublet and \Hb\ lines at the 
edge of the mosaic.

In a plane-parallel geometry, the ionization parameter describes the ratio of the ionizing flux at 
the inner surface of a cloud relative to the number density of hydrogen atoms. To assign an ionization 
parameter to the O32 values, we adopt the $Z = 0.05 \zsun$ relation in Figure 1 of \cite{Kewley2002},
corresponding to $12 + \log (O/H) = 7.6$, which is the closest model to the SB metallicity. 
The ionization parameter, $q$, describes the maximum speed at which the radiation field can drive an 
ionization front through a cloud complex. O32 ratios of $15, 10, 5$ indicate $q = 1300, 1000, 
{\rm and~} 600$ \kms, respectively.
It follows that the timescale for ionizing gas clouds, 
 $ \tau < 1.5 \times 10^5 {\rm ~yr~} (r / 100 {\rm ~pc}) (q / 600 {\rm ~km/s})^{-1}$
on the scale of the compact star clusters, is shorter than the cluster ages.
This result is consistent with the ionization front having radii of several kpc.

Comparison to other models often requires the dimensionless ionization parameter,
$U \equiv q /c$, where our  O32 values of $ 15, 10, 5$ correspond to  
$\log U = -2.36, -2.48, {\rm and} -2.70$.  One systematic uncertainty affecting
the interpretation of O32 ratios is model geometry. Spherical models, for example, give ionization 
parameters $ \approx 0.2$~dex higher than the plane-parallel geometry adopted.\footnote{
        We read the commonly used dimensionless form of the ionization parameter $U$ 
        off Figure 5 in \cite{Berg2016}, where the model geometry is spherical
        and compared to \cite{Kewley2002}.}
An additional systematic correction upwards is needed in the \heii - emitting region where
some oxygen has been further ionized to O$^{+3}$.  Relative to three-zone models, a four-zone 
model systematically yields higher $U$ by up to 0.5~dex \citep{Berg2021}. 
Since the ionization cone extends well above the region where the gas-phase
O/H ratio has been directly measured, we also considered the potential impact of 
metallicity gradients.
Metals in the central region of J1044+0353 have likely been diluted 
by the accretion which fueled the starburst, and the outflowing gas may have been enriched
by cooling of a hot wind \citep{Peng2023}.  At fixed O32, adopting the curve for a  higher 
metallicity in Figure 1 of \cite{Kewley2002} would further increase the estimated 
ionization parameter in the ionization cone.  Our argument that the O32 map reveals an ionization
cone therefore appears to be robust to these systematic uncertainties.

\subsubsection{Evidence for Anisotropic Lyman Continuum Escape}  \label{sec:lyc_esc}

The ionization cone defined by the anisotropic O32 emission in Figure~\ref{fig:O32_j1044}
{\clm 
provides new insight into the controversy over} the relationship between measured O32 ratios and \fesc
{\clm \citep{Faisst2016,Naidu2018}. } 
{\clm
Previous studies, see Figure 8a of \cite{Izotov2021} and 
Figure 10a of \cite{Bassett2019} for example,}
demonstrate that galaxies with high \fesc\ also have high O32 ratios, 
but that galaxies with high O32 ratios do not necessarily have {\it directly} detectable Lyman continuum.  
The scatter in the relationship between the O32 ratio and the directly measured Lyman 
continuum leakage has been interpreted as evidence that O32 is a mediocre predicter of 
\fesc\ for any individual galaxy \citep{Izotov2020a}. 
While their conclusion accurately describes leakage directly along our sightline to a galaxy, we
interpret the anisotropic O32 emission in Fig.~\ref{fig:O32_j1044} as evidence that
the global LyC leakage can be highly anisotropic. {\clm Our data therefore support the sketch in 
Figure 11 of \cite{Bassett2019} where  a galaxy with an ionization cone pointed away (toward) the 
observer produces high O32 (\fesc) but low \fesc\ (O32).}  Galaxy orientation may therefore
determine which strong W(\oiii) emitters have directly detected Lyman continuum emission
\citep{Tang2021b}. If anisotropic escape is common, we argue that O32 could be a more  robust measure of the 
global \fesc\ than the residual FUV continuum. 

{\clm Knowledge of the orientation and covering factor of density-bounded channels, in addition to 
measurements of gas-phase metallicity and ionization parameter, will likely be required in order to
accurately calculate the relationship between the O32 ratio and \fesc\ via photoionization modeling.}
Density bounded models produce large O32 ratios because the outer, low-ionization 
zone emitting \oii\ is truncated;  an observed O32 ratio can be produced with a lower ionization 
parameter than an ionization-bounded model would require \citep{Pellegrini2012,Jaskot2013}. 
An an example, we applied the density-bounded models shown in Figures 4 and 16 of \cite{Plat2019} to 
the ionization cone in Figure~\ref{fig:O32_j1044}. Where ${\rm O32}  \approx 15$, corresponding  to 
$F(\lambda 5007)/ F(3727) \approx 11$, those density bounded models indicate \fesc\ of 80\% and 
$\log {\rm U(R_S)} \approx -3.35$.  
{\clm Making such an inference from a homogeneous model, however, could be very misleading.  For example,
\cite{Bassett2019} fit the empirical relation between \fesc\ and O32 \citep{Faisst2016,Izotov2018b}
with a series of density-bounded, photoioniztion models.  Although they found a good fit by varying 
the \hi\ optical depth, the required metallicity and ionization parameter did not match the properties
of the galaxy sample. 
}
Since homogeneous, density bounded models also predict very weak \oi\ emission, 
the high \oi\ $\lambda 6300$ / \oiii\ $\lambda 5007$ flux ratio (O1O3) 
of Lyman continuum leakers presents another, perhaps 
related, puzzle  \citep{Plat2019}.  Two-zone models, where one one zone is density bounded, offer 
a possible solution \citep{Ramambason2020} which can be tested by resolving the escape channels.

The \lya\ profile supports the idea that LyC escape is highly anisotropic from J1044+0353. 
Radiative transfer models predict that LyC and \lya\ photons escape through the same holes \citep{Kakiichi2021}.
The {\it Hubble}  COS spectrum of SB reveals a double-peaked \lya\ emission-line profile; however,
as is often the case when the  spectroscopi aperture is much smaller than the solid angle of the \lya\ nebula,
the emission-line profile sits in a broad, dark H~I absorption trough \citep{Hu2023}.  Profiles of this 
type require a cloudy or multiphase ISM.  \citet{Hu2023} fit the \hi\ trough intensity along with
\hi\ damping wings, obtaining  an \hi\ column density of $\log N_{\hi} ({\rm cm}^{-2}) = 21.84 \pm 0.03$
over $91 \pm 3$ percent of starburst solid angle.  In other words, directly along our viewing angle,
the LyC leakage is  $ \le\ 9\pm3\%$.  
When $\fesc \le 0.10 $, \lya\ peak separations are typically greater than 300 \kms\ (\citet{Izotov2018b}, Figure~10;
\citet{Gazagnes2020}). 
The peak separation in the COS spectrum,  425 \kms, is therefore
consistent with little LyC leakage along the direct sightline,
and the asymmetry of the red peak is  also low, $A_f = 1.84 \pm 0.36$ \citep{Hu2023}.  
The \lya\ profile therefore supports the anisotropic geometry indicated in Figure~\ref{fig:O32_j1044}.

Based on the empirical relation \citep{Izotov2021}, we estimate that the Lyman continuum escape 
fraction exceeds 5\% where the O32 ratio (scaled to our definition in Section 4.2.1) exceeds 10
in Fig.~\ref{fig:O32_j1044}.  Perhaps most importantly, however, our observations strongly suggest
that the direction of the bipolar outflow powered by FEN dictates the direction of the ionization 
cone powered by SB, {\clm thus underlining the importance of the star formation history.}
{\clm In future work, we will present a larger sample of O32 maps, addressing the question of whether 
J1044+0353 is representative of \oiii-selected, dwarf galaxies, and then introduce O1O3 maps 
in order to test two-zone models and potentially measure \fesc\ indirectly.
}

\section{Gas Kinematics in the Starburst Region} \label{sec:gas_kinematics}

In this section we describe the gas kinematics in the starburst
region from multiple perspectives:  the ionized gas shells in Sec.~\ref{sec:ha_bubbles}, 
the blueshifted resonance absorption lines in Sec.~\ref{sec:uv_al}, and finally the broad 
emission-line wings in Sec.~\ref{sec:broad_wings} .  Viewed together, these data determine 
the launch radius and constrain the mass flux of a nascent wind. We discuss the required supernova rate in 
Section~\ref{sec:feedback_discussion} and the radiative losses from the nascent wind in 
Section~\ref{sec:cooling}.

\subsection{Starburst-driven Superbubbles} \label{sec:ha_bubbles}

In this section, we use the properties of the  prominent \Ha\ shells in Figure~\ref{fig:zoom_trio} 
to estimate the mechanical feedback from the young star clusters.  We predict the expansion 
velocties of the shells and then use the shell radii and velocities to quantify their power 
requirements. We  compare our result to the mechanical feedback predicted by population synthesis 
models in Section~\ref{sec:feedback_discussion}. 

Although there is more energy in the radiation field, supernova are the dominant feedback 
mechanism after the shells have emerged from the cloud cores and the Str\"{o}mgren radius 
has propagated through the shells.  Thermalization of the energy from clustered supernovae 
creates hot bubbles whose thermal pressure drives shock fronts into the ISM. The shells in 
Figure~\ref{fig:zoom_trio} consist of interstellar gas swept up by these shocks. We describe
the growth rate of these shells using the \cite{Weaver1977} model, replacing the stellar 
wind by a series of supernova explosions. In a roughly coeval stellar population, such as 
a compact star cluster,  the supernova rate will be nearly constant for $\sim   40$~Myr 
following the first supernova explosion because the slope of the initial mass function 
is almost perfectly offset by  the age -- lifetime relation of massive stars \citep{Leitherer1999}.

The shells are likely about 2 Myr old. This estimate is the difference between the upper limit on the 
starburst age, $4.01 \pm 1.13$~Myr \citep{Olivier2022}, and the 3~Myr delay for the first supernova 
explosion.\footnote{ 
             We note that this burst age was derived from population synthesis modeling based on 
             BPASS binary-star models.  Starburst~99 models (single stars) indicate a younger age, 
             $1.04 \pm  2.75$ ~Myr \citep{Olivier2022}, which is marginally too young to have 
             produced supernova explosions.}
For a constant rate of mechanical energy injection, the shell velocity is
\begin{eqnarray}
V = 100~ \kms\ 
     \bigg ( \frac{R}{ 340 {\rm  ~pc}} \bigg )
     \bigg (\frac{2 {\rm ~Myr}}{t} \bigg ).
\end{eqnarray}
In Figure~\ref{fig:zoom_trio},  the northwest shell extends 395~pc (1\farcs42) from the
star clusters in  Region~303, and the southwest shell has a radius of 341~pc (1\farcs12) 
with respect to the cluster in Region~304. At $t = 2$~Myr, the \citet{Weaver1977} model 
predicts expansion speeds of 116 \kms\ and 100 \kms, respectively, for the northwest and 
southwest shells.

Combining the equations for the evolution of the shell radius and velocity \citep{Weaver1977} 
gives us an expression for the power requirement:
\begin{eqnarray}
L_w = 1.85 \times 10^{41} \ergsec
     \bigg ( \frac{R}{ 340 {\rm  ~pc}}     \bigg)^{2} \times\ \\
     \nonumber
     \bigg ( \frac{V}{100~ {\rm km~s}^{-1}} \bigg)^{3} 
     \bigg ( \frac{n}{4 {\rm ~cm}^{-3}}     \bigg )
     \bigg ( \frac{\Omega}{4\pi}     \bigg ),
\label{eqn:lw} \end{eqnarray}
where $n$ is the hydrogen number density in the ambient medium. We take
the RMS density in the very-high ionization zone, see Section~\ref{sec:ne_rms},
as a fiducial value but discuss this choice further in Sec.~\ref{sec:feedback_discussion}.  
Following \citet{Geen2022}, we include the solid angle $\Omega$ of the shell because
the NW and SW shells emerge from narrow chimineys emanating from the Regions 303 and 304.
Inserting a solid angle $\Omega < 4 \pi$ in Equation~\ref{eqn:lw}  reduces the power
requirement linearly. We measure opening angles of approximately $45\deg$ and 
$75\deg$ for the northwest and southwest shells, respectively, corresponding to solid
angles of 0.478 and 1.298 steradians.  We conclude that the
growth of the northwest and southwest shells requires a mechanical power of
$L_w = 1.48 \times 10^{40}$ ergs s$^{-1}$ and $L_w = 1.92 \times 10^{40}$ ergs s$^{-1}$,
respectively, the equivalent of  900 and 1200 supernova explosions over the 2 Myr period.

On galactic scales, however, superbubbles elongate along the direction of the 
steepest density gradient \citep{MacLow1988}. Our density measurements, see  {\clm
Figure~\ref{fig:ne_map}}, indicate that the  steepest gradient is to the west 
of the starburst region, consistent with superbubble breakthrough in that direction.  
In the above argument, we ignored density gradients in the ambient medium because
the seeing-limited KCWI data cube does not resolve the ambient density profile 
$n(r)$ on the scale of the superbubble shells resolved by the {\it Hubble} imaging.
{\clm
A shell driven into a medium with an inverse square density profile does not decelerate,
so the predicted expansion speeds would increase to 192 and 166 \kms\ at 2 Myr.  
}
Direct measurements of the expansion velocities would clearly be valuable.  
{\clm
Expansion faster than the predicted value of 100 to 116 \kms\ would indicate a 
steep density gradient in the ambient medium. Whereas very low shell velocities,
or order 10 \kms, would longer timescales for star formation in the starburst region.
}

In the following two subsections, we describe  unequivocal kinematic evidence for strong 
outflows from the starburst. In the future, mapping infrared recombination lines with JWST would be
the cleanest way to directly measure shell exansion speeds, as NIRSpec could provide the 
high spatial and spectral resolution required.

\subsection{Ultraviolet Absorption-line Measurements of Outflow Velocity} \label{sec:uv_al}

Absorption in resonance lines is a powerful diagnostic of gaseous outflows and inflows
\citep{Martin2012,Heckman2015,Chisholm2017}. We inspected the {\it Hubble} COS UV spectrum of J1044+0353 
\citep{Berg2022,James2022}. The large red circle in Figure~\ref{fig:zoom_trio} shows that the 2\farcs5 (695 pc)
diameter aperture encompasses the \Ha\ shells. \cite{Xu2022} fit the Si II $\lambda 1190, 1193$, and 1304 
profiles with an interstellar component fixed at $v = 0$ and a blueshifted component at 
$v_0 = -52 \pm 12$ \kms. The width of the latter is 123 \kms\ FWHM, so the outflowing gas reaches 
speeds of  $v_0 + 0.5 {\rm FWHM} = 114 $\kms.
The maximum speeds therefore match the predicted  expansion speed of the northwest and southwest shells 
very well. In detail, however, this comparison is not straightforward because the spatial sampling of the 
absorption-line kinematics is determined by the distribution of UV light {\it behind} the shells. In the
simplest scenario, projection of the shell expansion velocity onto our sightline could explain why the line 
center appears at 52 \kms.
Previously, \citet{Xu2022} attributed the blueshifted absorption component to a galactic wind. 
Our analysis demonstrates that some of these {\it winds} are actually expanding shells 
\citep{Weaver1977} rather than the free-flowing winds described by \citep{CC85}.\footnote{
             To avoid confusion, we emphasize that J1044+0353 does have a global wind, which
             we map out in \citet{Peng2023}. This galactic outflow, however, lies entirely
             outside the COS spectrosopic aperture and is driven by an older stellar population. 
             Consistent with this conclusion, no high-ionization absorption is detected in  
             Si IV $\lambda \lambda 1393, 1402  $.}

The distinction between superbubbles and a galactic outflow matters because not all 
superbubbles blow out of the ISM, a topic we return to in Section~\ref{sec:cooling}; and, 
by associating the UV absorption with the shells, we determine the location of the absorbing 
gas along the line-of-sight.  The estimated mass flux scales linearly with radius, so the
detected shells imply mass fluxes orders of magnitude lower than circumgalactic shells would 
require.

The UV spectrum constrains the shell column density. We conservatively assume that Si$^{+}$ is 
the dominant ionization state of silicon atoms and ignore depletion onto dust grains. Then, using 
the gas-phase metallicity of 0.06\zsun, we obtain  $N({\rm H}) \approx 5.2 \times 10^{5}  (Z / 0.06 \zsun)^{-1}  
N({\rm Si~II}) $, which is $N(H) \approx 3.6 \times 10^{20}$~cm$^{-2}$ for the Si II column density
measured by \citet{Xu2022}.
The O I $\lambda 1302$ and C II $\lambda 1334$ line profiles are similar in shape  to the 
blueshifted Si II absorption troughs, and this demonstrates  that the shell contains some neutral gas.
The damped \lya\ absorption discussed in Section~\ref{sec:lyc_esc} has no significant Doppler
shift, but
\citep{Hu2023} identified a second H I component which is blueshifted 
and has a lower column density of  $\log N_{HI} ({\rm cm}^{-2}) = 20.08 \pm 0.44$.
We attribute this second \hi\ component to the expanding shells and emphasize that it has 
a very low covering fraction. This may reflect the small area of the shells in Figure~\ref{fig:zoom_trio} 
relative to the entire COS aperture.

For spherical shells, the mass flux is well defined at the shell radius:
\begin{eqnarray}
\dot{M}(r) = 1.8 \msunyr\ 
                 \bigg ( \frac{\Omega}{4\pi} \bigg )
                 \bigg ( \frac{N_H}{3.6 \times 10^{20} {\rm ~cm}^{-2}} \bigg ) \times\ ~~~\\
                  \nonumber
                  \bigg ( \frac{R}{340 {\rm ~pc}}                       \bigg ) 
                   \bigg ( \frac{V}{100 {\rm ~km/s} }      \bigg ),
\label{eqn:mdot} \end{eqnarray}
where $N_H$ represents the shell column density (ionized and neutral gas).  
If shells like the ones visible in Figure~\ref{fig:zoom_trio} were breaking through
the parent clouds in every direction, then the implied mass flux is $ \approx  1.8$\msunyr.
The outflowing mass flux of warm-ionized gas would be 12 times the aperture SFR from 
Table~\ref{tab:basic_properties}.  Making the correction for the  restricted solid angle 
of each shell, $\Omega \approx 0.48$ steradian and $\approx 1.298$ steradian, the two shells
combined transport  $\dot{M} \approx 0.28$\msunyr, a value that is roughly twice the (COS) 
aperture SFR of 0.15\msunyr.  These mass fluxes demonstrate significant transport of 
gas out of the starburst region.

\subsection{Two Origins of Broad Line Wings} \label{sec:broad_wings}

The starburst spectrum in Figure~\ref{fig:kcwi_spec} reveals broad wings on the \oiii\ and 
\Hb\ line profiles.  Further inspection of these line profiles across the KCWI mosaic reveals
that this high-velocity \oiii\ and \Hb\ emission comes from a spatially extended region 
centered on the starburst. In this section, we characterize the spatial extent, kinematics,
and luminosity of these broad line wings

\begin{figure*}[ht!]
\includegraphics[angle=0,width=7.0in]{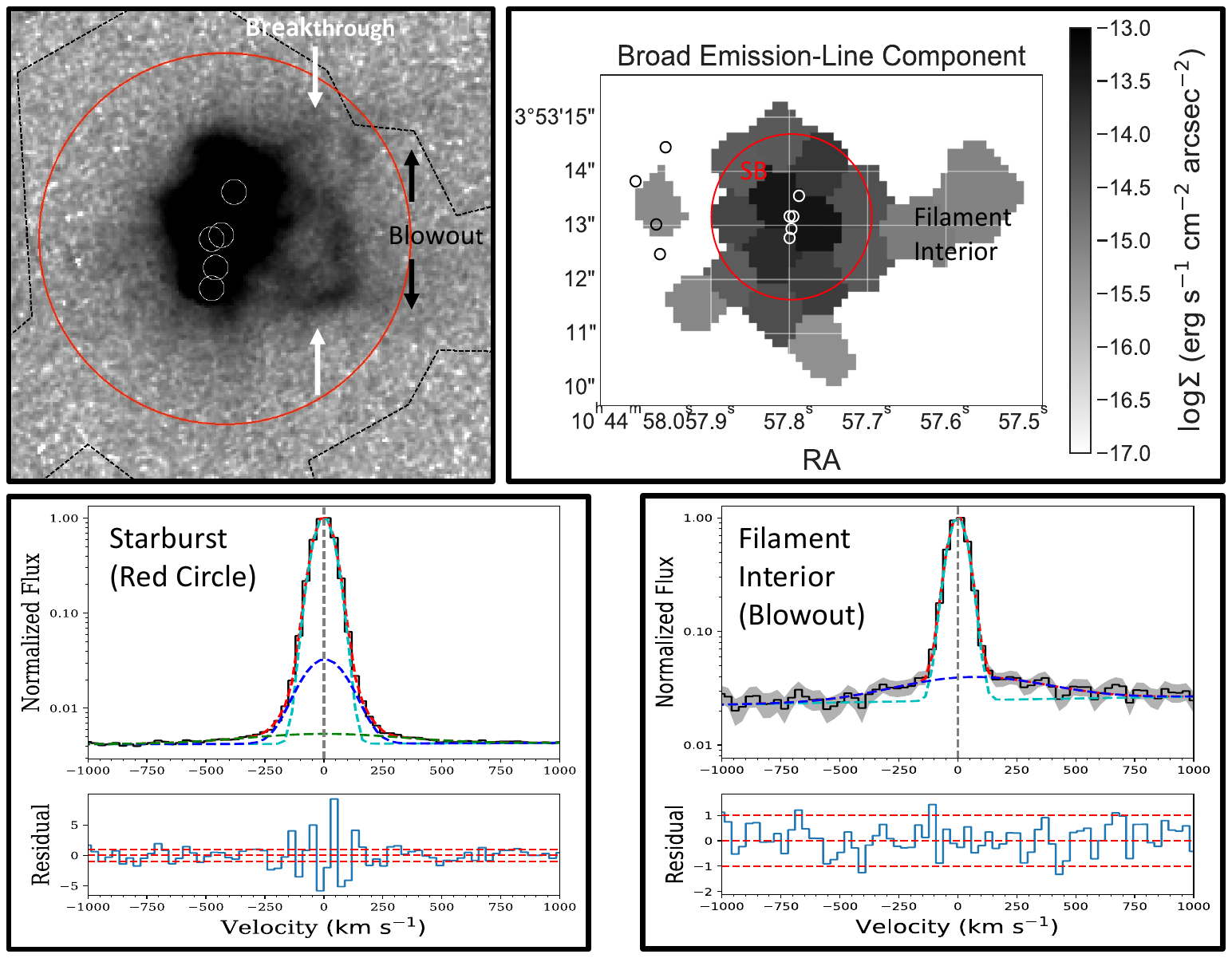}
\caption{Superbubble blowout in the J1044+0353 starburst. The radius of the red circle is 1\farcs53.
{\it Top Left:} {\it Hubble} \Ha\ image. Superbubbles {\it breakthrough} the central gas concentration, 
reaching 370~pc northwest and 340~pc southwest of the central star cluster. The open filaments beyond 
these closed loops are the signature of a superbubble {\it blowout}.
{\it Top Right:} KCWI map of the \oiii\ $\lambda 4959$ bins that pass the F-test for a second component.
The surface brightness is the sum of the fitted broad and very-broad components.
{\it Bottom Left:}  Starburst spectrum of \oiii\ $\lambda 4959$. A three-component fit is required:  
narrow component (97 \kms\ FWHM),  broad component (213 \kms\ FWHM, $ 3.17 \times 10^{-15}$ \flux), and
very-broad component (752 \kms\ FWHM, $4.5 \times 10^{-16}$ \flux).  The broad and very-broad components
are 5.7\%  and 0.8\%, respectively, of the total line flux.
{\it Bottom Right:} The \oiii\ $\lambda 4959$ spectrum extracted along the western filament requires
a two-component fit. The narrow component has a width of 93 \kms\ FWHM again. The width of the second 
component is 698 \kms\ FWHM, so we identify it as a physical extension of the very-broad component.
The flux of the very-broad component, $4.1 \times 10^{-17}$ \flux,  is relatively stronger in the filament,
where it accounts for 10\% of the  \oiii\ $\lambda 4959$ emission.
}
\label{fig:blowout_bel} \end{figure*}

To map these broad wings, we adaptively binned our \oiii\ $\lambda 4959$ line image 
to SNR $\approx 30$.  
We extracted a spectrum from 
each bin and first fit a  single Gaussian profile.  To determine whether a second component was 
detected, we then fit a double Gaussian profile and applied an F-test, as defined by Equation~1 
in \citet{Xu2022}.  If the calculated F value was larger than the theoretical value, a 
significance level $\alpha = 0.003 (3\sigma)$, then the extra component was considered necessary. 
The top-right panel of Figure~\ref{fig:blowout_bel} maps this high-velocity gas across 
J1044+0353.  We detected broad wings in and around the starburst region; these broad wings 
are absent in region FEN and the post-starburst region more generally.

The line profiles in the starburst region  require three Gaussian components, all at
the systemic velocity. The bottom-left panel of Figure~\ref{fig:blowout_bel} shows the \oiii\ fit.
Most of the line flux comes from the narrow component; this emission comes from 
photoionized clouds in the ISM.  
A broader component, 213 \kms\ FWHM, accounts for almost 6\% of the line flux toward the 
starburst. This {\it broad component} is likely emitted by the expanding shells. 
The  predicted shell velocities, 100 to 116 \kms\ in Section~\ref{sec:ha_bubbles}, 
correspond to linewidths of  200 to 232 \kms\  since we would see emission from  
their near (approaching) and far (receding) sides.

A two Gaussian fit to the starburst spectrum misses the very-broad wings. Their fitted width
is $752$ \kms\ FWHM, and we detect this component at Doppler shifts up to $\pm 1000$ \kms.  This 
{\it very-broad component} is not emitted by the expanding shells; its velocity is too large. 
In the upper right panel of Figure~\ref{fig:blowout_bel},  the spatial mapping of the line profiles 
offers important insight about the physical origin of  the {\it very-broad component}. To the 
north, south, and southeast, the very-broad wings are seen out to 2\farcs0 (560 pc), 3\farcs3 
(920 pc), and 3\farcs2 (880 pc), respectively.  This component extends further to the west, 
up to 4\farcs3  (1.20~kpc) from SB. Evidently,  one superbubble shell has already accelerated 
and fragmented, thereby launching a hot wind \citep{MacLow1988,MacLow1989}. In support of this 
conclusion, the direction of the western arm coincides with the steepest density gradient and the 
elongation of the superbubble shells. Following \citet{MacLow1989}, who showed that blowout
occurs at roughly three times the fiducial timescale $t_D \equiv H^{5/3} \rho_0^{1/3} L_w^{-1/3}$, 
we estimate a minimum age of 
\begin{eqnarray}
t_{blow} \approx 3 t_{D} \approx 3.7~{\rm ~Myr} 
    \bigg ( \frac{H}{200 {\rm ~pc}}  \bigg )^{5/3} \times\ ~~~ \\
    \nonumber
     \bigg ( \frac{10^{40} {\rm ~erg s}^{-1}}{L_w}  \bigg )^{1/3}
      \bigg ( \frac{n}{4 {\rm ~cm}^{-3}}  \bigg )^{1/3}.
\label{eqn:tblow} \end{eqnarray}
Seeing-limited data does not resolve the density field on the spatial scale of the parent 
gas clouds. We have inserted a fiducial value of  200~pc for the  e-folding scale, 
$H$, of the  density gradient; this value represents  the distance between Region~304 and
Region~301.
{\clm
This argument predicts that the superbubble shells are not only breaking through their parent gas
clouds but are also close to blowout.
}

The top left panel of Figure~\ref{fig:blowout_bel}  draws attention to two fainter
\Ha\ filaments. These largely radial filaments extend beyond the superbubble shells, and  
we suggest they are remnants of ruptured superbubble shells, relics of superbubble blowout. 
{\clm
The  KCWI data cube provides kinematic information on this larger spatial scale, and 
we show the region where the  \oiii\ emission line requires a second velocity component
in the top right panel of Figure~\ref{fig:blowout_bel}. } The odd extension to
the west of SB falls within the two radial \Ha\ filaments. Whereas the starburst spectrum,
shown in the lower left panel, has three velocity components, the broad component is absent
bewteen in this western arm; yet the very-broad component persists. The bottom-right panel 
of Figure~\ref{fig:blowout_bel} shows our fit to a spectrum extracted along this western arm. 
Whereas the very-broad component comprises just 0.8\% 
of the total line flux in the starburst region, it grows to 10\% of the total \oiii\ flux in 
the western arm. {\clm We therefore suggest that superbubble blowout, a nascent wind, produces
the very-broad component.} The absence of the broad component in the western arm is consistent 
with the superbubble shells being outside the spectroscopic aperture.  We estimate that the solid 
angle of the outflow is at least 10 \% of $4\pi$ steradian. {\clm We discuss the implications
for launching a galactic wind further in Section~\ref{sec:feedback_discussion}.}


\section{Discussion} \label{sec:discussion}

In this section we compare our measurements of the radiative and mechanical
feedback from the J1044+0353 starburst to expectations from stellar population
synthesis modeling. We model the temporal evolution of the ionizing continuum and 
the mechanical feedback using the Binary Population and Spectral Synthesis v2.1 (BPASS) 
models \citep{Eldridge2017}. 
In these models mass transfer between binary stars mixes a larger fraction 
of the zero-age stellar envelope into the core, producing longer lived and hotter stars 
than do traditional models.  Including the evolution of massive-star binaries in evolutionary 
synthesis models therefore boosts the \heplus-ionizing luminosity relative to 
population synthesis models based on single stars \citep{Eldridge2009, Eldridge2012}. 
We adopt these models because they produce the hardest ionizing continua among currently 
available population synthesis models; but the systematic uncertainties are substantial,
and the soft X-ray emission is not included.

In Section~\ref{sec:xion}, we revisit the spectral hardness problem in J1044+0353 based
on our new measurements of ionizing luminosity from Section~\ref{sec:luminosities} and 
discuss whether stellar populations can produce the ionizing photons.
We discuss the implications of the supernova-driven shells for the starburst properties
in Section~\ref{sec:feedback_discussion}. 
In Section~\ref{sec:cooling}, we discuss radiative cooling of the hot wind where the
superbubble has blown out of the starburst region.

\subsection{Ionizing Photon Production Efficiency} \label{sec:xion}

The ionizing photon production efficiency, \xion,  compares the luminosity of the ionizing continuum
to the non-ionizing UV continuum.  We expect high  \xion\ in J1044+0353 because \xion\ shows a strong 
positive correlation with increasing W(\oiii) \citep{Chevallard2018}. In Section~\ref{sec:recombination},
we  measured the ionizing photon luminosity  at two energies, 13.6 eV and 54.4 eV. This {\it absorbed} 
photon luminosity is a lower limit on the total luminosity because a fraction, \fesc, 
of the ionizing radiation presumably escapes from the galaxy.

Dividing the ionizing luminosity by the extinction-corrected, non-ionizing UV luminosity defines the production
efficiency for ionizing photons, 
$
\xion(X) \equiv Q(X) / L^{UV}_{\nu},
$
where X denotes either the H-ionizing or the \heplus-ionizing continuum. 
This definition of \xion\ implicitly includes the contribution of nebular continuum as well as massive 
stars. Defined this way, \xion\ is the parameter needed to determine the ionization rate of intergalactic 
gas from the star formation rate density of the universe and knowledge of \fesc\ \citep{Bouwens2016}.
We measured UV magnitudes for both the Starburst and the entire galaxy in Sec.~\ref{sec:hubble_images},
In terms of the absolute UV magnitude,  this expression becomes
\begin{eqnarray}
\xion(X) = 2.31 \times 10^{24} {\rm ~Hz~erg}^{-1} \bigg( \frac{{\rm Q(X)}}{10^{53} 
      {\rm ~s}^{-1}} \bigg ) \times\ ~~~\\
      \nonumber
10^{0.4 [M_{UV} + 20]}.
\label{eqn:xionH} 
\end{eqnarray}
Applying Eqn.~\ref{eqn:xionH} to the entire galaxy, $M_{UV} = -16.68$ from Table~\ref{tab:qphot},
we find $\log \xion ({\rm Hz~erg}^{-1}) = 25.5$. If we limit measurement to the UV luminosity of the 
SB, the ionizing photon production efficiencies increase to $\log \xion ({\rm Hz~erg}^{-1}) = 25.7$ and 
$\log \xionhe ({\rm Hz~erg}^{-1}) = 23.6$. 
For the harder photons that can ionize \heplus, Eqn.~\ref{eqn:xionH} yields 
$\log \xionhe ({\rm Hz~erg}^{-1}) = 23.5$. 
In contrast to the growth of W(\oiii) with increasing aperture size, the \xion\ and \xionhe\ measurements
are 1.25 to 1.5 times larger in the Starburst aperture than for the galaxy as a whole.

\begin{figure}[h]
\includegraphics[angle=0,width=3.5in]{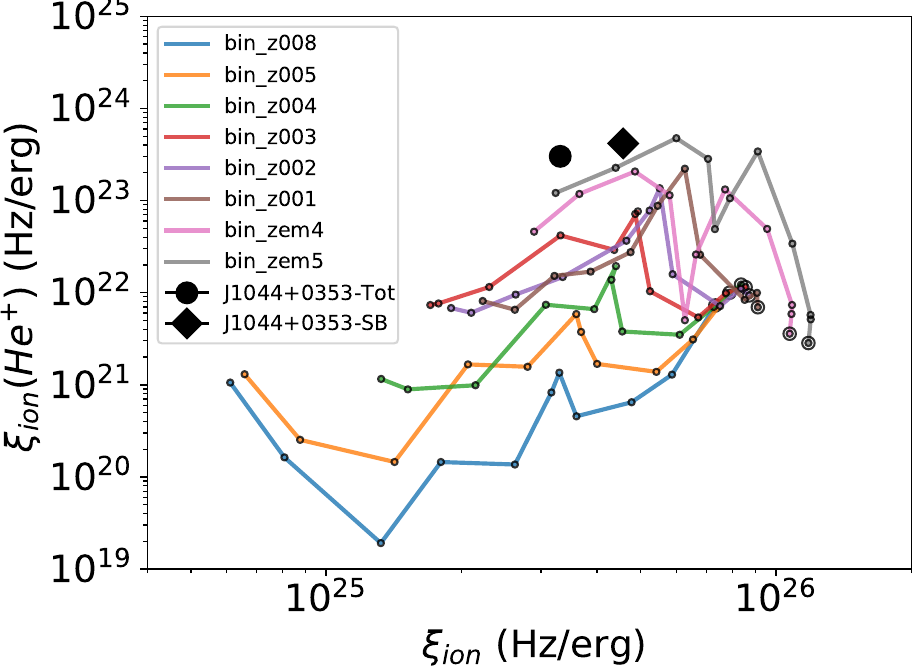}
\caption{Ionizing photon production efficiency in J1044+0353 (black symbols) compared BPASS 
tracks for a coeval population of binary stars  \citep[BPASSv2.1]{Eldridge2017,Xiao2018}. 
The stellar metallicities shown range from $12 + \log (O/H) = $ 5.6 (zem5) to 8.52 (z008).  
Age increases from right ($\log t (yr) = 6.0$) to left ($\log t (yr) = 7.0$) in steps of 0.1 dex.
The  model at the most appropriate metallicity,  bin\_z001, reaches a maximum \xion(\heplus), 
$ 2.2 \times 10^{23}$~Hz~erg$^{-1}$, near our measured value but at a younger  age of 2.5~Myr. 
At the measured age of 4 Myr, the model is full dex below our measured value. This spectral
hardness problem is alleviated by the models with extremely metal-poor stellar populations.
}
\label{fig:xion_xion_binary} \end{figure}

When working with population synthesis models, the ionizing photon luminosities are normalized 
directly by the stellar continuum. We call this quantity $\xion_*$ here to distinguish it from
\xion. \citet{Olivier2022} estimated $\log \xion_* ({\rm Hz~erg}^{-1})= 25.8 \pm 0.9$  by modeling 
the FUV spectrum of J1044+0353, so  the difference  is just $\xion_* - \xion \approx 0.1$~dex for 
J1044+0353. We computed \xion$_*$ and \xionhe$_*$ for the BPASS 
models by integrating the number of photons at energies exceeding 13.6~eV and 54.4~eV, respectively, 
and normalizing by the flux density, $F_{\nu}$ at 1500 \AA\ in the model spectra. In these
models, the production efficiency of H ionizing photons decreases continuously as the  stellar 
population ages, but the temporal evolution of the \heplus\ ionizing spectrum depends on stellar 
metallicity. The tracks for the lowest metallicity models in Figure~\ref{fig:xion_xion_binary} show
\xion(\heplus) rising for several Myr before beginning a slow decline. The brown track, model 
bin\_z001,  corresponds to 12 + log(O/H) = 7.61 based on Table 2 of \cite{Xiao2018}. This track
is the closest match to the gas-phase metallicity of J1044+0353,  
$12 + \log (O/H) = 7.45 \pm 0.03$ \citep{Berg2022}.

Figure~\ref{fig:xion_xion_binary} compares our measurements of the ionizing photon
efficiencies to the BPASS tracks. {\clm Model bin\_z001 is the nearest match to J1044+0353.}
Our estimated \xion\ for J1044+0353 falls within the model range between ages of 3 to 6 Myr.  
Corrections for escaping Lyman continuum  shift our measurements to higher \xion, and therefore 
require younger ages; but the correction is less than 0.1 dex in \xion\ for $\fesc \le 0.21$.  
The production efficiency of \heplus\ ionizing photons peaks at an age of 2.5 Myr, where 
$\xion_*(\heplus)$, just 70\% of  our measurement. At the favored age of 4 Myr, the models are 
1.1 dex lower than our measurement. {\clm Correction for the stellar contribution to the 
\heii\ $\lambda 4686$ luminosity, $4^{+16}_{-4} \% $ in Section~\ref{sec:he_second_component}, 
reduces \xionhe\ but leaves a significant excess of hard photons relative to model bin\_z001.}

{\clm
We show two lower metallicity BPASS models in Figure~\ref{fig:xion_xion_binary}  because
previous work suggests that the the excess of \heplus-ionizing photons in some starburst galaxies is 
related to their low metallicity \citep{Chevallard2018,Schaerer2019,Senchyna2020,Marques-Chaves2022,Sander2022}. }
The bin\_zem5 model with $12 + \log (O/H) = 5.60$ meets \xion(\heplus) requirement at  $ t = 2.5, 
5.0, {\rm ~and~} 6.3$~Myr.  At an age of 4 Myr there is no significant spectral hardness problem,  and 
a LyC escape fraction of 16\% would be enough to  increase \xion\ to the model's value.  However,
such extreme stellar metallicity is not consistent with the stellar wind and photospheric features previously
detected in the FUV spectra of J1044+0353; they indicate a higher stellar metallicity   
$Z_* = 0.14 \pm 0.04$\zsun \citep{Olivier2022}.  Super-solar \aFe\ abundance ratios have also been  
proposed to explain the hard radiation fields of young galaxies because iron largely 
determines the far-UV opacity of stars \citep{Steidel2018,Topping2020, Hirtenstein2021,Sander2022}. Interestingly, 
however, although the photospheric S V $\lambda 1502$ absorption in the J1044+0353 FUV spectrum is 
weaker than in a comparison sample, the Fe V absorption in the same spectrum is actually stronger than 
the S V absorption \citep{Olivier2022}. It follows that the J1044+0353 starburst does not show
the hypothesized super-solar \aFe\ abuandance ratio.  We conclude that BPASS models cannot
explain the high \xion(\heplus) in J1044+0353 and confirm the previously recognized
spectral hardness problem \citep{Olivier2022}. {\clm In Section~\ref{sec:he_second_component}, we presented
evidence for VMS in J1044+0353. Including VMS with low-metallicity in stellar population synthesis 
models is expected to reduce this tension; whether it eliminates the discprepancy, however, remains unknown
at this time.}

\subsection{Supernova Feedback} \label{sec:feedback_discussion}

The superbubble shells presented in Section~\ref{sec:gas_kinematics} demonstrate
strong mechanical feedback from the starburst.  In this Section, {\clm we discuss
the significance of very young stellar populations producing supernovae}
and then compare the number of required supernova to stellar population synthesis models.

\begin{figure}[h]
\includegraphics[angle=0,width=3.75in]{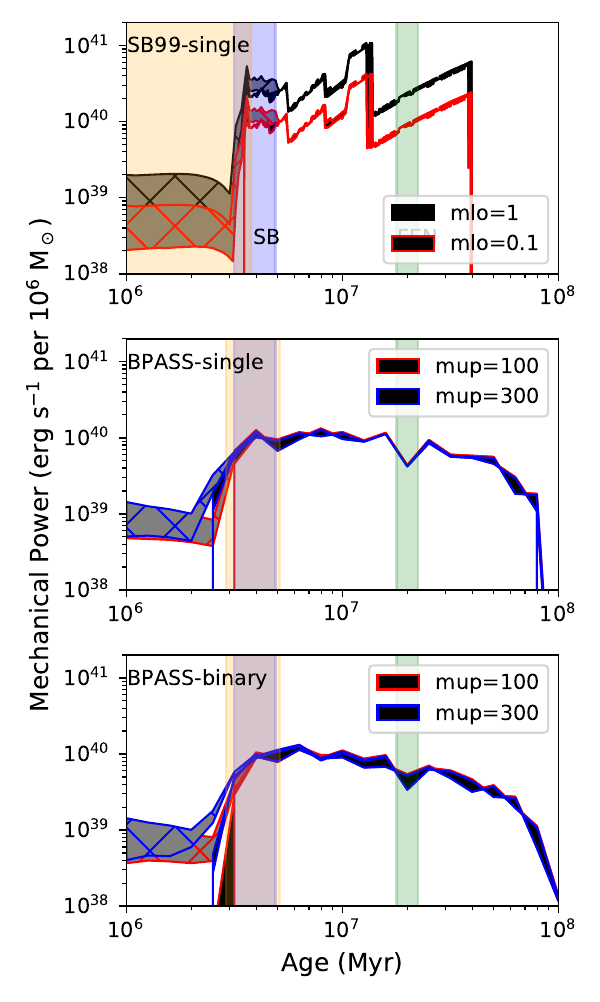}
\caption{
Mechanical feedback from population synthesis modeling of
a single-age stellar population.
Supernova feedback and total feedback including stellar winds are
indicated by black and gray shading, respectively; the shaded range 
represents stellar metallicities ranging from roughly 5\% to 20\% \zsun.
The ages of the starburst (SB) and the far-eastern clusters (FEN) 
are indicated by the vertical shading.  The SB region in J1044+0353
has just begun to produce supernovae, whereas the clusters
near the base of the bipolar outflow have already produced many supernovae.
}
\label{fig:sb99_bpass_lw} \end{figure}

Figure~\ref{fig:sb99_bpass_lw} illustrates the temporal evolution of the  mechanical feedback 
from several widely-applied stellar population synthesis models.  The delay for the 
first supernova is two to three Myr, reflecting a weak dependence on the upper mass limit.
Starburst~99 models fitted to the UV spectrum require a starburst age of only $1.04 \pm 2.75$ Myr 
\citep{Olivier2022}, and these models produce the first supernova at an age of 3.3 Myr. 
Figure~\ref{fig:sb99_bpass_lw} demonstrates that the BPASS binary-star models alleviate this tension.
With the same upper mass limit, the evolutionary tracks used by BPASS produce supernova 
slightly sooner,  at 2.6~Myr; but, the fitted starburst age increases to $4.03 \pm 0.85$ 
Myr \citep{Olivier2022}.  In Section~\ref{sec:ha_bubbles}, we adopted a fiducial 
starburst age of 5 Myr and a delay of 3 Myr for the first explosion to obtain
2 Myr for shell growth. {\clm Population synthesis models make assumptions about 
which massive stars produce supernovae.  The minimum mass for core collapse determines
a low-mass cutoff around 8-10\msun. Modeling when stellar core collapse produces a 
supernova explosion continues to be a challenging problem, however \citep{Fryer2023,Heger2023}, 
as core collapse in stars with initial masses above 40\msun\ is not always expected to 
produce a supernova \citep{Smartt2015,Sukhbold2020}. Hence, it is not clear how the
presence of very massive stars would
change the supernova rate per unit stellar mass.}
Feedback observations can therefore help address these substantial 
uncertainties regarding the supernova rate from young, metal-poor  stellar populations.

Starburst~99 population synthesis models run with a Kroupa initial mass function
yield a fiducial mechanical power of $L_{SN} = 
1.7 \pm 0.1 \times 10^{40}$\ergsec\ per $10^6 \msun\ $ of stars.
Using the  starburst masses from either Table~\ref{tab:hst_phot} or Table~\ref{tab:basic_properties},
the available power is $L_{SN} = 2.3 \times 10^{40}$ ergs~s$^{-1}$
or $L_{SN} = 2.6 \times 10^{40}$ ergs~s$^{-1}$, respectively, where we take the 
difference as an indications of the systematic uncertainty in the stellar mass estimate.
Then using the stellar masses from Table~\ref{tab:hst_phot}, we  estimate that
Region 303, at the base of the northwest shell, produced 
$L_{SN} = 5.9 \times 10^{39}$ergs~s$^{-1}$;  and Region 304 produced $L_{SN} =
5.8 \times  \times 10^{39}$~ergs~s$^{-1}$. These values are only  39\% and 30\% of our
estimated power requirements for the NW and SW shells, respectively, in Section~\ref{sec:ha_bubbles}.
{\clm 
While a stellar population that produces more mechanical power per unit mass can eliminate this
discrepancy, the uncertainties in the required power allow other solutions.}

Those estimated power requirements scale linearly with the ambient density, for example; see
Equation~\ref{eqn:lw}.  In Section~\ref{sec:ha_bubbles}, we adopted the RMS density of 
the warm gas as the fiducial ambient density.  Below, in  Section~\ref{sec:cooling},
we argue for emergence of a hot phase, a three-phase ISM. The ambient density, and hence
the required mechanical power, would be lower if the hot phase is the volume-filling phase.
Conductive evaporation from the superbubble shells produces 
an interior temperture \citep{Weaver1977} 
\begin{eqnarray}
T_b = 1.30 \times 10^7 {\rm ~K} 
            \bigg (  \frac{L_w}{5 \times 10^{39} {\rm ~erg~s^{-1}}}  \bigg )
             \bigg (  \frac{n_0}{4 {\rm ~cm^{-3}}}  \bigg ) \times\ ~~~ \\
             \nonumber
             \bigg (   \frac{2 {\rm ~Myr}}{t} \bigg )
              \kappa_0^{-2/7},
\label{eqn:t_bubble} \end{eqnarray}
where $\kappa_0 \approx 1$.  A hot phase at a fiducial temperture of 
$T_h = 1 \times 10^7$~K has of  density of $n_{0,h} \approx 0.37 {\rm ~cm}^{-3} (10^7 {\rm ~K} / T_h)$ 
in pressure equilibrium with in the warm ($2 \times 10^4$ K) clouds, where 
the electron density derived from the \oii - doublet ratio is 185~cm$^{-3}$.
An  ambient density of 0.4~cm$^{-3}$ would reduce the required mechanical power
by a factor of ten, and a normal IMF for Regions 303 and 304 would
produce 4 and 3  times the required mechanical power. A low ambient density 
would also reduce the blowout timescale given by  Equation~\ref{eqn:tblow}, just
 1.7 Myr after the first supernova for a density $n_0 = 0.4$ cm$^{-3}$.

\subsection{Origin of the Very Broad \oiii\ Wings} \label{sec:cooling}

The \oiii\ luminosity of the  {\it broad} velocity component exceeds the rate of 
mechanical energy deposition in J1044+0353.  Our schematic places the origin of this 
broad velocity component in the superbubble shells, consistent with photoionization 
heating this component.  The superbubble blowout identified
in Figure~\ref{fig:blowout_bel} is noteworthy because it connects the {\it very-broad} 
velocity component to the process of launching a galactic wind. The \oiii\ luminosity of
this very-broad component, which is much lower than that of the broad component, plausibly
comes from cold, outflowing clouds which exchange mass, momentum, and energy with a hot
wind. In Section~\ref{sec:beta}, we constrain the mass-loading of that  hot wind and 
discuss whether it cools catastrophically. In Section~\ref{sec:cooling} we further discuss
the luminosity of the very-broad component.

\subsubsection{Mass Loading of the Hot Wind} \label{sec:beta}

In this section we model the linewidth of the very-broad component using a single-phase 
wind model that includes radiative cooling. \citet{Thompson2016} present a picture 
where entrained clouds seed a cooling instability. Adding mass to the hot wind lowers the wind
temperature.  As a hot wind cools through the temperature range where the cooling coefficient 
increases, roughly $2 \times 10^7 > T ({\rm K}) > 2 \times 10^5$ 
\citep[chapter 34][]{Draine2011}, runaway radiative cooling can stall the wind. 

The mass-loading parameter $\beta $ describes the mass flux of the hot wind in units of the SFR, 
$\beta \equiv \dot{M}_{hot} / \dot{M}_{*}$. The \citet{Thompson2016}  model defines the minimum amount of mass loading, 
$\beta_{cool}$,  required for the cooling radius, where the cooling time drops below the 
expansion time, to move inward. Mass loading sufficient to move the cooling radius inwards 
all the way to the starburst radius  defines a critical mass loading parameter, 
$\beta_{crit}$. Writing Equation 10 of \citet{Thompson2016}  in units appropriate to J1044+0353, 
we have
\begin{eqnarray}
\beta_{crit}(r_{cool} = R_{SB}) \approx 3.1 \alpha^{0.730} 
 \bigg [\frac{\zeta}{0.587} \bigg ]^{0.730} \times\ ~~~ \\
 \nonumber
  \bigg [\frac{\Omega}{4 \pi} 
         \frac{R_{SB}}{100 {\rm ~pc}}
          \frac{0.15 \msunyr}{\dot{M}_*}
           \bigg ]^{0.270},
\label{eqn:bcrit} \end{eqnarray}
where we have included a  parameter $\zeta$ to make the aperture SFR give the same 
mechanical power as the aperture mass.\footnote{
     \citet{Thompson2016} adopt a constant supernova rate 
     equivalent to 1 supernova per unit SFR every 100 years.  For a constant SFR, however, the 
     supernova rate rises steadily for the first 40 Myr, so we cannot apply their fiducial 
     scaling to a very young starburst. In addition, for consistency with the rest of this paper, 
     where we have parameterized the mechanical feedback in terms of the burst mass, we adopted
     a mechanical power,  $L_{SN} =  1.76 \times 10^{40} {\rm~erg~s}^{-1}~  ( \zeta / 0.587 ) 
     ( \dot{M}_* / 1.0 \msunyr )$, where $\dot{M}_*$ is the SFR. }

The adiabatic wind model originally described by \citet{CC85} predicts a maximum wind velocity
\begin{eqnarray}
v_{term} =  980~ \kms\  \alpha^{0.5} \beta^{-0.5},
 \label{eqn:vterm} \end{eqnarray}
where the parameter $\alpha$ describes the energy deposition rate relative to the supernova 
rate. Energy loss through radiative cooling will decelerats the hot wind. Equation~14 of  
\citet{Thompson2016} predicts that the terminal velocity falls to
\begin{eqnarray}
v_{term} = 440~ \kms\ 
     \bigg ( \frac{\zeta }{0.587}           \bigg )^{0.135}
     \bigg ( \frac{ \alpha \xi}{\Omega_{4\pi}}           \bigg )^{0.135} \times\ ~~~ \\
     \nonumber
     \bigg (  \frac{\dot{M}_{*}}{0.15 ~ \msunyr}        \bigg )^{0.135} 
     \bigg (   \frac{100 ~ {\rm pc}}{R_{SB}}        \bigg )^{0.135},
\label{eqn:vcrit_bcrit} \end{eqnarray}
in the limit $\beta = \beta_{crit}$.  The wind fails if $\beta > \beta_{crit}$,
so this velocity describes the minimum outflow velocity. The metallicity parameter
$\xi \approx 1$ describes a solar metallicity wind, an appropriate value in a dwarf
galaxy \citep{Martin2002}.

In a scenario where the ram pressure of the wind accelerates interstellar gas clouds, 
the velocity of the clouds cannot exceed the speed of the hot wind \citep{Murray2005}. 
It follows that the terminal velocity of the hot wind limits the maximum Doppler shift 
of the \oiii\ line wings. {\clm Hence we can use observational constraints on $\beta$ and 
the terminal wind speed in order to discuss  whether strong cooling causes the nascent 
wind to fail.}

In Section~\ref{sec:uv_al}, we estimated the  mass flux in the two \Ha\ shells. When
these shells undergo blowout, warm gas mass will be entrained in the hot wind at a rate 
of 0.28 \msunyr. {\clm Normalized by the SFR, we estimate $\beta = 0.7 - 1.9$.\footnote{
     We obtained an average SFR by dividing the stellar masses in Table~\ref{tab:hst_phot} 
     by the corresponding cluster age. Relative to regions~303 and 304, we obtain a mass loading 
     parameter $\beta \sim 1.1$. Alternatively, normalizing by the SFR of the entire starburst 
     region gives  $\beta \sim 0.7 - 0.8$. We find a larger value, $\beta \approx 1.9$, 
     when we use the aperture SFR from Table~\ref{tab:basic_properties}. The differences 
     between these normalizations define the uncertainty in our estimate of $\beta$. }
}
These modest mass-loading factors are near unity, the fiducial value in the adiabatic wind 
model. In addition, the faint wings of the \oiii\ emission line reach $\pm 1000$ \kms, a good 
match to the terminal wind velocity predicted by Equation~\ref{eqn:vterm}. We therefore draw 
two conclusions: (1) the wind has a hot phase, and (2) some portion of that wind is
nearly adiabatic. 

The typical cloud speed is significantly lower than 1000 \kms. The half-width at half 
maximum intensity of the very-broad component suggests a speed closer to 380 \kms.
We combined Equations~\ref{eqn:vcrit_bcrit} and~\ref{eqn:bcrit} and found  that reducing
$v_{term} $ to 380 \kms\ requires $\beta_{crit} = 3, ~5, {\rm ~and~} 10$ for 
$\alpha = 0.5, 1, {\rm ~and~} 2$, respectively.  {\clm These estimates for the critical
mass loading parameter are independent of the solid angle of the outflow. Since they
are significantly larger than the estimated value of $\beta$, we argue that the 
nascent wind will not undergo catastrophic cooling.}

\begin{figure}[h]
\includegraphics[angle=0,width=3.5in, trim = 20 0 0 0]{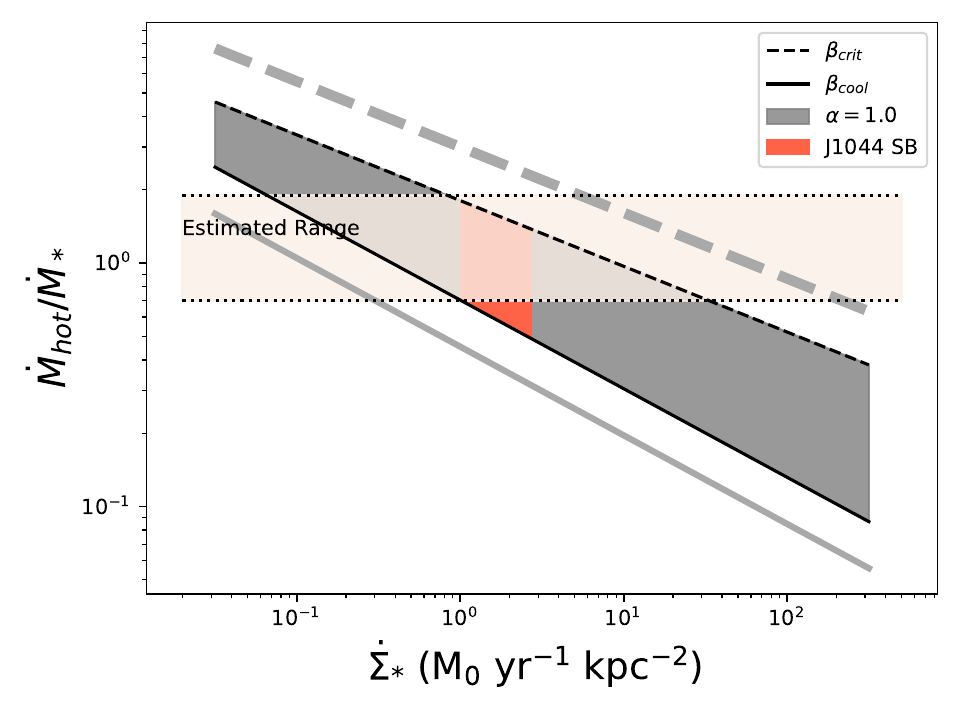}
\caption{The minimum mass-loading required to get cooling anywhere in the wind 
(solid lines) and the maximum mass-loading beyond which the entire wind cools (dashed lines), as 
given by Equations 7 and 10 in \citet{Thompson2016}. Models that lie below the $\beta_{crit}$ line 
identify viable models for launching a wind; winds with $\beta > \beta_{crit}$ cool catastrophically.
The negative slope indicates that at fixed $\beta = \dot{M}_{hot}/\dot{M}_{*}$ there exists a
critial SFR surface density above which the gas cools too strongly at blowout to form a steady wind.
Our fiducial wind model adopts a starburst radius $R_*= 100$~pc, solid angle $\Omega = 4 \pi$, 
and $\alpha \equiv \dot{E}_{hot} / L_{SN} = 1$. The intersection of this model with the SFR surface 
density of J1044+0353 is shaded orange. Our $\beta$ measurement, shown by the horizontal bar, has 
significant overlap with the orange region where the radiative energy losses are not catastrophic. 
The thick dashed (solid) line illustrates how raising (lowering) the supernova rate by settting
$\alpha = 2$ ($\alpha = 0.5$) raises (lowers)  $\beta_{crit}$ ($\beta_{ccool}$).}
\label{fig:cooling_wind}
\end{figure}

{\clm
Figure~\ref{fig:cooling_wind} illustrates how $\beta_{crit}$ and $\beta_{cool}$ fall 
with increasing SFR surface density. We mark the SFR surface density of the
J1044+0353 starburst.\footnote{
      The starburst region of J1044+0353 has $\dot{\Sigma}_* \approx 2.39^{+0.32}_{-0.48}$ \msunyr\ 
      kpc$^{-2}$ based on the aperture SFR and half-light radius from \citet{Berg2022} and 
      \citet{Xu2022}, respectively. We estimate a lower value value, $\dot{\Sigma}_* \approx 1.0 $ 
      \msunyr\ kpc$^{-2}$, from the starburst model in Table~\ref{tab:hst_phot} and the area of the 
      red circle in Figure~\ref{fig:kcwi_fov_duo}.  }
The shaded region illustrates the range of $\beta$  where our fiducial wind model cools radiatively but 
without catastrophic energy losses. The wide horizontal bar denotes our estimate of 
the mass loading parameter. The overlap between the viable models, the horizontal bar,
and the starburst SFR/Area selects the locus of viable models. Catastrophic
cooling, in contrast, would require $\dot{\Sigma}_* > 30$ \msunyr\ kpc$^{-2}$ for the
chosen parameters. Raising $\alpha$ from 1 to 2, which is equivalent to doubling the 
number of supernova per unit SFR, would allow SFR surface densities up to 
$\dot{\Sigma}_* <  200$ \msunyr\ kpc$^{-2}$ to generate  viable winds. 
}

%
%


\subsubsection{Radiative Cooling of the Hot Wind} \label{sec:cooling}

Radiative cooling plays a key role in cloud survival, and perhaps growth, as hydrodynamical 
instabilities otherwise shred the entrained clouds \citep{Gronke2018}.  The single-phase model 
discussed in Section~\ref{sec:beta}  describe the wind with a single temperature at a given radius,
so it does not include the mixing layers between clouds and the hot wind. These mixing layers
likely dominate the total cooling \citep{Fielding2022}. The \oiii\ luminosity of the very-broad 
component provides insight about the cooling mechanism.

The very-broad component  has a total, reddening-corrected luminosity  
$L(\oiii\ \lambda 4959, 5007)  \approx 1.1 \times 10^{39}$ \ergsec. 
We estimated $ L_{SN} \approx 2.64 \times 10^{40}$ \ergsec\ 
from the aperture mass in Table~\ref{tab:basic_properties}. 
Since energy is put into the hot wind at a rate $\dot{E}_{hot} = \alpha L_{SN}$, it 
follows that the luminosity emitted in the very-broad wings of the \oiii\ doublet is 
$0.04 \alpha^{-1} \dot{E}_{hot}$. As a single-phase wind must cool from an initial temperature
$T \sim  few \times 10^7$~K, first free-free radiation and then emission from highly
ionized metals including O~VII, and O~VI lines, carry away energy. The gas temperature must 
decrease  to a $few \times10^5$~K before \oiii\ becomes an important coolant, so the 
the thermal energy has been reduced by a factor of order 100 before \oiii\ becomes an
important coolant. We can confirm this back-of-the-envelope argument by applying
Equation~36 of \citet{Thompson2016}. The result suggests that the energy radiated (primarily
in UV and optical lines) is $L_{UV/O} \sim 7.0 \times 10^{38} (\dot{M} / 1 \msunyr)$ ergs~s$^{-1}$
as the wind cools from $10^5$ to $10^4$~K. Taking $\dot{M} = \beta \dot{M}_* < 5 \dot{M}_* $, 
the model  predicts a maximum luminosity $ L_{cool}(10^4-10^5 ~{\rm K}) \sim 
5.3 \times 10^{38}$~ergs~s$^{-1}$, roughly 2\% of $L_{SN}$ as anticipated. 

Only a portion of this $L_{cool}$ comes out in the \oiii\ lines, so the model requires
$L_{\oiii}) < L_{cool}$. {\clm We examined the \citep{Ploeckinger2020} models to 
gain insight into the fraction of $L_{cool}$ emitted in \oiii\ $\lambda 5007$ line. 
In their models the cooling gas is exposed to an interstellar radiation field.  
As the gas cools from $10^{5.5}$~K to $10^4$~K, about 1\% of the cooling luminosity 
is emitted in \oiii\ $\lambda 5007$.} Yet the measured \oiii\ luminosity of the very-broad 
component is comparable to $L_{cool}$ from the single-phase wind model. In future work, 
we will further described this luminosity problem using a larger sample of galaxies
with very-broad emission lines. For now, we suggest that the discrepancy draws attention
to the limitations of single-phase wind models and underlines the importance of multiphase 
wind modeling. Our results  motivate calculating the fraction $L_{SN}$ that emerges as 
$L_{\oiii}$ under realistic conditions in outflows.

Some models for cooling winds predict strong \heii\ emission \citep{Danehkar2021}, so
we revisted our interpretation of the $\pm 400$ \kms\ \heii\ line wings
in Section~\ref{sec:he_second_component}. The \heii\  line has comparable 
S/N ratio to \oiii\ $\lambda 4363$, so we examined the broad component of
\oiii\ $\lambda 4363$ in order to understand the detectability of broad \heii\ components.  
The auroral \oiii\ line has a broad base. A joint fit with the nebular \oiii\ returns
the broad component described in Section~\ref{sec:broad_wings}. It is therefore remarkable
that a joint fit including \heii\ $\lambda 4686$ produces negative residuals where
a broad \heii\ component might be expected.  We therefore conclude that
the  gas which emits the broad \Hb\ and \oiii\ $\lambda 4959$ components does not
contribute to the \heii\ luminosity.

\section{Summary \& Conclusions} \label{sec:summary}

In this paper, we presented integral field spectroscopy of the reionization-era analog, J1044+0353, 
an \oiii - selected galaxy with strong \heii\ line emission. We combined several KCWI pointings into 
a mosaic cube designed to cover the  galactic Str\"{o}mgren sphere.  At several thousand independent 
locations, we analyzed the intensity and profile shape of emission lines in the 3700 - 5600 \AA\ 
bandpass and described the global gas kinematics, density gradients, filling factor,  and ionization structure. 
We leveraged  high-resolution of {\it Hubble} images to better understand the morphology
of massive star clusters and superbubbles in the  starburst region.  We compared the supernova
rates and the ionizing photon productions rates, \xionh\ and \xionhe, predicted by
stellar population synthesis models to our measurements of these quantities in J1044+0353.
We summarize the main results here and draw attention to what they teach us about 
galaxies widely believed to have reionized the universe.

\paragraph{The Spectral Hardness Problem}

{\clm
Previous studies have concluded that the hardness of the ionizing spectrum in starburst galaxies 
increases as metallicity decreases  
\citep{Senchyna2017,Chevallard2018,Senchyna2020,Marques-Chaves2022,Schaerer2022a,Flury2022},
but whether low-metallicity stars can supply all the hard ionizing photons remains an active topic of discussion
\citep{Senchyna2017,Berg2018,Ravindranath2019,Saxena2020a,Olivier2022,Harish2023}. 
Our KCWI observation revealed that the \heii\ $\lambda 4686$ emission from J1044+0353 comes from 
an ellipsoid with semi-minor and semi-major axes of 610 and 730 pc, respectively, centered on
a pair of compact star clusters, Region 303.  Most of the \heii\ luminosity has a nebular origin, 
\heplusplus\ recombination,  based on the narrow core of the \heii\ line profile and the spatial
coincidence of the \heii\ and \ariv\ emission regions. Based on the spatial elongation of this very-high
ionization zone, which follows the locus defined by five massive stars clusters, we suggest that
the source of the \heplus-ionizing photons is not a single source, such as an AGN.
Normalized by the UV luminosity of the starburst, the ionizing photon luminosity, \qheplus,
corresponds to an ionizing photon efficiency  $\xionhe = 23.53$~Hz~erg$^{-1}$, which 
exceeds the values produced by BPASS models of an appropriate metallicity and age.
Through a close examination of the \heii\ line profile, which is broader than the \ariv\ line profile,
we estimated that $4^{+16}_{-4}\%  $ of the \heii\ line luminosity comes from a second component, not
of nebular origin. The most likely source of this {\it broader} component is stellar wind emission. 
Very massive  ($> 100$ \msun), hydrogen-core burning stars are the most likely source of \heii\ wind 
emission from a stellar population only a few Myr old \citep{Grafener2008,Wofford2023}. The relatively 
slow stellar wind, i.e. the width of the second component is just $\pm 200$ \kms\ from line center,  
then indicates the presence of stars with significant rotation, and thus likely stellar metallicity
as low the measured gas-phase metallicity \citep{Meynet2002,Grafener2015}. This small stellar contribution 
to the \heii\ luminosity does not significantly mitigate the spectral hardness problem, but it does
underline the pressing need for accurate theoretical modeling of the \heplus-ionizing continuum from
low-metallicity star clusters.
}

\paragraph{The Supernova Delay Problem} 

Simultaneous fits of several line ratios are typically used to determine \fesc; this information 
breaks degeneracies between ionization-bounded and density-bounded models \citep{Plat2019} or 
between density-bounded channels and density-bounded galaxies \citep{Ramambason2020}. We previously
mapped the gas-phase oxygen abundance and interstellar reddening in J1044+0353 \citep{Peng2023}. 
In this paper, we identified a density-bounded ionization cone based on the anisotropic structure 
seen an F(\oiii\ $\lambda \lambda 4959,5007) / F(\oii \lambda \lambda 3726,29)$ flux ratio map. 
The O32 ratio is high in the starburst region, declines with radius at most position angles, but
remains high throughout a density-bounded ionization cone  extending north-northeast. {\clm A key
part of this argument relied on \oiii\ detections at large separations, 1.5 - 2.5 kpc, from 
the starburst region where no \oii\ emission was detected, yielding a lower limit on O32.
}
Along the line of sight to J1044+0353, only a small fraction of the starburst is 
covered by low column density \hi\ \citep{Hu2023}. Hence, we argue that the LyC leakage 
from J1044+0353 is primarily in the direction of the ionization cone.  This anisotropic
escape results from the ionizing luminosity, produced in the starburst region, encountering 
the cavity excavated by a bipolar outflow launched from region FEN. The older stellar age of 
FEN makes this region a strong source of mechanical feedback even though its current production
rate of ionizing photons is negligible compared to the compact starburst. At least 
in J1044+0353, the star formation history solves the supernova delay problem.

\paragraph{The Catastrophic Cooling Problem}

{\clm
In numerical simulations, galactic winds form through superbubble breakthrough and 
blowout \citep{MacLow1988,MacLow1989}.}
Remarkably, we show that superbubble breakthrough and blowout imprint distinct 
signatures on the starburst spectrum. Toward the starburst region, the \oiii\ and \Hb\ 
line profiles show a broad (213 \kms\ FWHM) and a very broad (752 \kms\ FWHM) components. 
We mapped these components individually.  The broad component has the higher luminosity, 
is spatially extended, and is centered on the starburst region.  The two most prominent 
superbubbles within this region extend 350 and  400~pc, respectively, from the compact
star clusters {\clm labeled Region 304 and 303.} Growing these bubbles requires shell velocities  
$v_{SW} \approx 100 (t/2 {\rm ~Myr})^{-1}$ \kms\ and $v_{NW} \approx 120 (t/2 {\rm ~Myr})^{-1}$ 
\kms. The starburst age from \citet{Olivier2022} therefore produces  a good match 
between the shell velocity and the blueshift of UV absorption lines \citep{Xu2022}. 
We therefore suggest that the outflowing gas identified by \cite{Xu2022} resides in 
a superbubble shell rather than a multiphase galactic wind; the actual bipolar wind
form J1044+0353 is outside their spectroscopic aperture. The superbubble dynamics 
constrain the supernova rate in star clusters believed to have formed some very massive stars,
a population where empirical constraints on supernova rates are needed \citep{Fryer2023}.
A filamentary arm extending 1 kpc to 
the west of the starburst appears only in the lower intensity, very-broad component.  
Close examination of the {\it Hubble } \Ha\ image reveals faint \Ha\ emission on each side 
of this very-high velocity gas. These features are consistent with superbubble blowout,
and the velocity of the very-broad component matches expectations for the velocity of
a free-flowing, hot wind \citep{CC85}. These observations suggest a picture where
the young starburst is starting to launch a galactic wind. 
A comparison to the 
\citet{Thompson2016} wind model demonstrated that the radiative cooling from 
this nascent wind is not catastrophic. The \oiii\ luminosity of the wind is higher than 
single-phase wind models can explain, however; and the discrepancy may provide a useful 
target for multi-phase wind models. These results aid the interpretation of the 
emission-line wings detected in 25-40\% of redshift $3 < z < 9$ galaxies  
\citep{Carniani2023,Xu2023}.

\vskip\the\baselineskip

The galaxy population believed to drive cosmic reionization had low masses, 
very low abundances of heavy elements, and high star-formation efficiencies 
\citep{Ouchi2009,Wise2009,Bouwens2015}.  JWST is finding  a large population 
of such galaxies via their strong \oiii\ emission  \citep{Oesch2023,Tang2023,Matthee2023}.
In this paper, we have demonstrated that the star formation history across a local analog, 
J1044+0353, plays a critical role in forming pathways for LyC escape. 
The LyC photons generated by the very young starburst encounter an ISM whose structure has already 
been modified by a galactic wind launched 10-20 Myr prior from a post-starburst region. 
These results support the idea that galaxy orientation determines 
which strong W(\oiii) have directly detected LyC \citep{Tang2021b}. We suggest that due to
anisotropic LyC escape, spatial mapping of the O32 ratio may turn out to be a more 
accurate indicator of  LyC escape than  direct detection of the Lyman cotinuum, which
requires an optically thin channel to be aligned with our sightline. 

{\clm 
Our schematic picture of J1044+0353 is qualitatively similar to the scenario seen 
in full-physics simulations of reionization-era galaxies; the SFR is highly variable 
in time, and LyC escape is anisotropic \citep{Martin-Alvarez2023,Witten2024}.
Bursty star formation histories at redshift $z \sim 10-13$ also offer an explanation 
\citep{Munoz2023,Sun2023a,Sun2023b} for the surprising luminosity excess 
observed \citep{Naidu2022,Finkelstein2023,Casey2023}. The rapid build up of galaxies
during the reionization era produces many galaxy -- galaxy mergers \citep{Witten2024}.
In contrast, it is not obvious that local analogs selected to resemble high-redshift 
galaxies based on their spectral properties would share this environmental property.
Indeed, the stellar mass of J1044+0353 is well below the limits of pair surveys 
\citep{Stierwalt2017,Besla2018}, so the probability of a dwarf -- dwarf merger 
in the $M_* \sim 10^6$  \msun\ to $10^7$ \msun\ range is not constrained observationally.
}
J1044+0353 has neither a luminous companion nor any known tidal tails, i.e. features
that would confirm an interaction or merger hypothesis. We note, however, the unexplained 
presence of a band of high extinction, elongated perpendicular to a line drawn from SB 
to  FEN \citep{Peng2023};  the coincident string of older star clusters there resembles Mrk 709
\citep{Kimbro2021}. Hence we suggest that a dwarf -- dwarf merger triggered the recent
star formation activity in J1044+0353.
{\clm
Mergers of dwarf galaxies have been observed to produce more widely distributed star 
formation than mergers of  massive galaxies, which funnel gas into a central starburst 
\citep{Privon2017}, and we would expect the first passage to produce a starburst in 
both  dwarfs \citep{Kimbro2021}. A merger scenario  predicts a positive radial age 
gradient in the starburst region, similar to the one measured in 30~Doradus 
\citep{Brandl1996,Cignoni2015} and some interstellar gas flung out to 
large distances \citep{Pearson2016,Pearson2018} and possibly detectable via
21-cm mapping.
}

\acknowledgments
We thank George Privon, Zirui Chen, Mordecai Mac-Low, Drummond Fielding, and Norman Murray 
for discussions that improved the paper.  Pierre Thibodeaux developed the adapative binning
code used here.
The data presented herein were obtained at the W. M. Keck Observatory, which is operated as a 
scientific partnership among the California Institute of Technology, the University of California and 
the National Aeronautics and Space Administration. The Observatory was made possible by the generous 
financial support of the W. M. Keck Foundation. The authors wish to recognize and acknowledge the very 
significant cultural role and reverence that the summit of Maunakea has always had within the indigenous 
Hawaiian community.  We are most fortunate to have the opportunity to conduct observations from this mountain. 
This work was completed at the Aspen Center for Physics, which is supported by National Science Foundation 
grant PHY-2210452. This work was supported by the National Science Foundation through AST-1817125.

%

\vspace{5mm}
\facilities{WMKO(KCWI), HST(ACS,WFC3)}


\software{
{\sc Astropy} \citep{Astropy2013},
{\sc BPASS} \citep{Eldridge2017},
{\sc Cloudy} \citep{Ferland2013},
{\sc CWITools} \citep{OSullivan2020},
the {Infrared Science Archive} ({\url https://irsa.ipac.caltech.edu/frontpage/}),
{\sc KDERP} ({\url https://github.com/Keck-DataReductionPipelines/KcwiDRP}),
{\sc LMFIT} \citep{Newville2021},
{\sc MVT-binning} ({\url https://github.com/pierrethx/MVT-binning } )
{\sc PyNeb} \citep{Luridiana2015},
and
{\sc SExtractor} \citep{Bertin1996}
}
          




\newpage

\begin{table*}[h!]
 \caption{Basic Properties of J0144+0353}
  \begin{tabular}{llll}
   \hline
Property        &  Derived                                 & Unit                 & Reference \\
    \hline
     \hline   
W(\oiii)            & $1040\pm20$                        & \AA                       & NSA Catalog \\
M$_R - 5 \log h$    & -14.95                        & mag                       & NSA Catalog\tablenotemark{a} \\
M$_{FUV} - 5 \log h$ & -15.40                        & mag                       & NSA Catalog\tablenotemark{a} \\
E(B-V) (Galactic)  &  0.0366                        & mag                  &     \citet{Schlafly2011}   \\
A$_{\rm V}$        & $0.20^{+0.10}_{-0.10}$    &   mag                 &  \cite{Peng2023}            \\
%
%
SFR$_{Tot}$ (SED)    &$0.26^{+0.07}_{-0.07}$               &   \msun\ yr$^{-1}$    &      \citet{Berg2022}\tablenotemark{b}            \\
SFR$_{Ap }$ (SED)    &$0.13^{+0.02}_{-0.03}$               &   \msun\ yr$^{-1}$    &      \citet{Berg2022}\tablenotemark{b}               \\
$\log {\rm M}_{*,Tot}$ (SED) & $6.8^{+0.41}_{-0.26}$            &   \msun               &        \citet{Berg2022}\tablenotemark{b}             \\
$\log {\rm M}_{*,Ap}$  (SED) & $6.13^{+0.23}_{-0.09}$           &   \msun               &        \citet{Berg2022}\tablenotemark{b}             \\
$r_{50} (UV)$           & 0.38                                &   arcseconds             &  \citet{Xu2022}            \\
SFR/Area           &$0.6^{+0.17}_{-0.14}$                &   \msun\ yr$^{-1}$ kpc$^{-2}$  &  \citet{Xu2022}            \\
log Specific SFR  &$-7.39^{+0.43}_{-0.28}$               &           yr$^{-1}$   &           \citet{Xu2022}            \\
$v_{circ}$        & $15.6^{+3.0}_{-3.7}$                 & \kms                 &           \citet{Xu2022}            \\
$v_{out}$ (UV, LOS)    & $52 \pm 13$                          &  \kms               &      \citet{Xu2022}            \\
$v_{out,fwhm}$ (UV)    & $123 \pm 24$                          &  \kms               &      \citet{Xu2022}            \\
v$_{out}$ ([O III])& +50 to -40           &  \kms             &  \cite{Peng2023}            \\
$T_e$   ([O III])  & $19,200\pm200$               &   K              &     \citet{Berg2021}             \\
$T_e$   ([O II])   & $19,100\pm1500$               &   K              &     \citet{Berg2021}             \\
12 + $\log$ (O/H)  &$7.45\pm0.03$               &                  &    \citet{Berg2022}              \\
$n_e$   ([S II])   &$200\pm40$                 &   cm$^{-3}$       &    \citet{Berg2022}              \\
$n_e$   (C III])   & $ < 9000$                 &   cm$^{-3}$       &    \citet{Berg2021}              \\
$\log \xion$        &$25.8\pm 0.9$               &  Hz~erg$^{-1}$        &  \citet{Olivier2022}                \\
SB Age (UV Spectrum, BPASS) & $ 4.03 \pm 0.85 $                       &   Myr               &  \citet{Olivier2022}                \\
SB Age (UV Spectrum, SB~99) & $ 1.26 \pm 2.36 $                       &   Myr               &  \citet{Olivier2022}                \\
SB  Age (Balmer EW) &$6.480 \pm 0.005$           &  Myr               &  \cite{Peng2023}                \\
FEN Age (Balmer EW)&$ 7.30\pm 0.05$             & Myr                &  \cite{Peng2023}  \\
NEN/MEN Age (Balmer EW)&$7.2 \pm 0.1$               & Myr               &  \cite{Peng2023} \\
\hline
      \end{tabular}  \label{tab:basic_properties}
       \tablenotetext{a}{Scaled to the distance adopted in this paper gives M$_{FUV} = -16.23$
         M$_R = -15.78$.} 
       \tablenotetext{b}{Assumes a luminosity distance of 55 Mpc. Scale by 1.15 to match the distance adopted
         throughout this paper.}
       \end{table*}


\begin{deluxetable*}{llllllll}[h!]
\tablecaption{Starburst Photometry and SED Fitting}
\tablehead{
\colhead{} &
\colhead{SB} &
\colhead{SW-shell} &
\colhead{NW-shell} &
\colhead{301} &
\colhead{302} &
\colhead{303} &
\colhead{304} 
}
\startdata
F125LP (mJy) & $0.1181$  & $0.0079  $   & $0.00506  $                          & 0.00324    & 0.00324    & 0.0312      & 0.0359    \\
F150LP (mJy) & $0.1362$  & $0.0089  $   & $0.0061 $                            & 0.00334    & 0.00334    & 0.0379      & 0.0436    \\
F165LP (mJy) & $0.1334$  & $0.0100  $   & $0.0065 $                            & 0.00333    & 0.00333    & 0.0362      & 0.0397    \\
F336W  (mJy) & $0.1729$  & $0.0109  $   & $0.0064 $                            & 0.00635    & 0.00635    & 0.0532      & 0.055     \\
F438W  (mJy) & $0.174$  & $0.0119  $   & $0.0073 $                            & 0.00636    & 0.00636    & 0.0521      & 0.0527    \\
F606W  (mJy) & $0.322$  & $0.0173  $   & $0.0126 $                            & 0.0132     & 0.0132     & 0.101415    & 0.0999    \\
F665N  (mJy) & $1.5615$   & $0.077  $    & $0.0631 $                            & 0.067      & 0.067      & 0.502318    & 0.48167   \\
F814W  (mJy) & $0.1105$  & $0.0086  $   & $0.0056 $                            & 0.00408    & 0.00408    & 0.0314      & 0.0308    \\
\hline                                                                                                                       
$M$ ($10^{5}$ \msun)\tablenotemark{a} & $11.78^{+0.34}_{-0.20}$ & $1.31^{+0.29}_{-0.19}$ & $0.41^{+0.01}_{-0.01}$ & $0.52^{+0.04}_{-0.02}$ & $1.04^{+0.06}_{-0.05}$ & $3.67^{+0.18}_{-0.17}$ & $3.44^{+0.15}_{-0.15}$ \\
$t_{ssp}$ (Myr)\tablenotemark{a} & $2.98^{+0.01}_{-0.01}$ & $5.02^{+0.33}_{-0.22}$ & $3.03^{+0.01}_{-0.01}$ & $2.95^{+0.02}_{-0.03}$ & $2.97^{+0.02}_{-0.03}$ & $2.97^{+0.02}_{-0.02}$ & $2.98^{+0.01}_{-0.02}$ \\
$\tau_eff$\tablenotemark{a} & $0.228^{+0.007}_{-0.007}$  & $0.248^{+0.032}_{-0.026}$ & $0.181^{+0.004}_{-0.007}$ & $0.337^{+0.012}_{-0.011}$ & $0.249^{+0.010}_{-0.011}$ & $0.254^{+0.011}_{-0.011}$ & $0.213^{+0.010}_{-0.010}$ \\
\hline
$M$ ($10^{5}$ \msun)\tablenotemark{b} & $11.43^{+0.62}_{-0.50}$ & $1.22^{+0.15}_{-0.16}$ & $0.42^{+0.02}_{-0.02}$ & $0.54^{+0.03}_{-0.03}$ & $1.07^{+0.06}_{-0.05}$ & $3.86^{+0.20}_{-0.20}$ & $3.52^{+0.15}_{-0.15}$ \\
$t_{ssp}$ (Myr)\tablenotemark{b} & $2.99^{+0.01}_{-0.02}$ & $4.88^{+0.18}_{-0.17}$ & $3.03^{+0.01}_{-0.01}$ & $2.95^{+0.02}_{-0.03}$ & $2.96^{+0.02}_{-0.03}$ & $2.96^{+0.02}_{-0.03}$ & $2.98^{+0.01}_{-0.02}$ \\
$\tau_eff$\tablenotemark{b} & $0.159^{+0.012}_{-0.015}$ & $0.240^{+0.019}_{-0.025}$ & $0.181^{+0.017}_{-0.013}$ & $0.348^{+0.012}_{-0.012}$ & $0.252^{+0.011}_{-0.012}$ & $0.261^{+0.012}_{-0.011}$ & $0.217^{+0.010}_{-0.011}$ \\
\hline
$M$ ($10^{5}$ \msun)\tablenotemark{c} & $11.74^{+0.63}_{-0.43}$ & $2.09^{+0.29}_{-0.15}$ & $0.78^{+0.19}_{-0.19}$ & $0.56^{+0.03}_{-0.03}$ & $1.12^{+0.05}_{-0.05}$ & $4.07^{+0.16}_{-0.17}$ & $3.80^{+0.15}_{-0.17}$ \\
$t_{max}$ (Myr)\tablenotemark{c} & $3.37^{+0.06}_{-0.08}$ & $21.88^{+3.83}_{-4.73}$ & $10.23^{+2.95}_{-2.98}$ & $3.24^{+0.05}_{-0.08}$ & $3.24^{+0.05}_{-0.05}$ & $3.24^{+0.05}_{-0.05}$ & $3.24^{+0.05}_{-0.05}$ \\
$\tau_eff$\tablenotemark{c} & $0.156^{+0.011}_{-0.014}$ & $0.220^{+0.012}_{-0.012}$ & $0.198^{+0.023}_{-0.025}$ & $0.290^{+0.013}_{-0.013}$ & $0.201^{+0.012}_{-0.012}$ & $0.208^{+0.012}_{-0.011}$ & $0.164^{+0.012}_{-0.013}$ \\
\hline
$M$ ($10^{5}$ \msun)\tablenotemark{d} & $12.01^{+0.75}_{-0.53}$ & $1.94^{+0.32}_{-0.29}$ & $0.74^{+0.18}_{-0.14}$ & $0.58^{+0.03}_{-0.03}$ & $1.15^{+0.05}_{-0.05}$ & $4.13^{+0.16}_{-0.17}$ & $3.86^{+0.18}_{-0.17}$ \\
$t_{max}$ (Myr)\tablenotemark{d} & $3.39^{+0.06}_{-0.11}$ & $17.34^{+2.80}_{-2.75}$ & $9.40^{+2.32}_{-2.74}$ & $3.23^{+0.07}_{-0.07}$ & $3.24^{+0.05}_{-0.05}$ & $3.24^{+0.05}_{-0.05}$ & $3.26^{+0.05}_{-0.05}$ \\
$\tau_eff$\tablenotemark{d} & $0.159^{+0.012}_{-0.016}$  & $0.318^{+0.052}_{-0.047}$ & $0.195^{+0.040}_{-0.025}$ & $0.294^{+0.015}_{-0.013}$ & $0.204^{+0.013}_{-0.012}$ & $0.211^{+0.012}_{-0.011}$ & $0.168^{+0.013}_{-0.012}$ \\
\enddata
\tablenotetext{a}{
Beagle fits using single-age stellar population and smc reddening. Age defined as the SSP age.
}
\tablenotetext{b}{
Beagle fits using single-age stellar population and CF00 reddening.
}
\tablenotetext{c}{
Beagle fits using const SFH and smc reddening. Age defined by the maximum stellar age.
}
\tablenotetext{d}{
Beagle fits using const SFH and CF00 reddening.
}
\tablenotetext{e}{
The photometric errors are dominated by systematic errors which are estimated to be 5\%.
}
\tablenotetext{f}{Comparison of the parameters for the four fitted models illustrate their systematic uncertainties. The ionization parameter $\log(U)$, stellar metallicity $\log(Z/Z_\odot)$, and gas-phase metallicity $\log(Z_{gas}/Z_\odot)$ were fixed at -1.75,  -1.296, and -1.296. A Chabrier IMF from 0.1 to 100 $\msun$ was adopted. Departures from these values would increase the systemic error. The statistical errors given were obtained by the 16th and 84th percentile of the posterior distribution of the parameters in comparison with the median (50th percentile) value.}
\label{tab:hst_phot}
\end{deluxetable*}



\begin{deluxetable*}{lll}[h!]
\tablecaption{KCWI Measurements of Radiative Feedback}
\tablehead{
\colhead{Measurement\tablenotemark{b,c}}      &
\colhead{Unit} &
\colhead{J104457+035313} 
}
\startdata
                   & \rc(\hplus) Aperture\tablenotemark{a}                       &                  \\
\hline
\rc(\hplus)        &       pc          & 1990    \\
\rx(\hplus)               &       pc          & $>3340$    \\
F(\Hb)                   &  \flux            & $4.600 \pm 0.002 \times 10^{-14}$    \\
F(\oii\ $\lambda 3726$)                       &  \flux            & $ 93.80\pm xx \times 10^{-16}$    \\
F(\oii\ $\lambda 3729$)                       &  \flux            & $ 122.49\pm xx \times 10^{-16}$    \\
L'(\Hb)                   &  ergs s$^{-1}$   & $ 2.54 \pm 0.15 \times 10^{40}$   \\
Q(H)                      &  s$^{-1}$        & $ 5.51\pm 0.22 \times 10^{52}$   \\
\nrmsq               & ${\rm cm^{-3}}$  & 0.69                     \\
$n_e (\oii)$      & $ {\rm cm}^{-3}$ & $161^{+7}_{-10}$                    \\
\fv (\oii)                          &    \nodata                & $1.8 \times 10^{-5}$                         \\
M$_{FUV}^{Tot}$                       & AB Mag           & $-16.68 \pm 0.17 $         \\
$\log \xionh_{Tot}$             &  Hz erg$^{-1}$         & $25.43^{+0.10}_{-0.13}$                          \\
$\log \xion(\heplus)_{Tot}$     &  Hz erg$^{-1}$         & $23.39^{+0.11}_{-0.14}$                           \\
%
\hline                                                                
\hline                                                                
& \rc(\heplusplus) Aperture\tablenotemark{a}                       &                  \\
\hline                                                                
\rc(\heplusplus) & pc                     & 610         \\
\rx(\heplusplus) & pc                     & 730          \\
W(\oiii)                        &  (\AA)             & 1230                      \\
F(\Hb)           &  \flux            & $3.752 \pm 0.002 \times 10^{-14}$    \\
F(\heii\ $\lambda 4686$) &  \flux         & $7.24 \pm 0.03 \times 10^{-16}$           \\
F(\oii $\lambda 3726$)                       &  \flux            & $ 76.08\pm xx \times 10^{-16}$    \\
F(\oii $\lambda 3729$)                       &  \flux            & $ 97.68\pm xx \times 10^{-16}$    \\
F(\oiii $\lambda 4959$)                       &  \flux            & $ 600.6\pm 0.1 \times 10^{-16}$    \\
F(\ariv\ $\lambda 4740$)                       &  \flux            & $ 6.47\pm xx \times 10^{-16}$    \\
F(\ariv\ $\lambda 4711$)                       &  \flux            & $ 7.75\pm xx \times 10^{-16}$    \\
L'(\heii\ $\lambda 4686$) & ergs s$^{-1}$ & $4.0 \pm 0.2 \times 10^{38}$   \\
Q(\heplus)                &     s$^{-1}$  & $ 4.89 \pm 0.47 \times 10^{50}$   \\
\nrmsq (\heii)    & ${\rm cm^{-3}}$  & 0.50                      \\
\nrmsq (\hi)   & ${\rm cm^{-3}}$  & 3.7                      \\
$n_e (\oii)$       & $ {\rm cm}^{-3}$ & $185^{+9}_{-11}$                    \\
\fv (\oii)                     &  \nodata                  & $4 \times 10^{-4}$                         \\
$n_e (\ariv))$ & ${\rm cm}^{-3}$  & $1440^{+44}_{-128}$                       \\
\fvh (\ariv)                    &  \nodata                  & $1.2 \times 10^{-7}$                         \\
M$_{FUV}^{SB}$               & AB Mag           & $-16.32 \pm 0.17$          \\
$\log \xionh_{SB}$           &  Hz erg$^{-1}$         & $25.58^{+0.098}_{-0.13} $                         \\
$\log \xion(\heplus)_{SB}$    &  Hz erg$^{-1}$        & $23.53^{+0.11}_{-0.14} $                          \\
\hline                                                                
\hline                                                                
            &\rc(\heplus) Aperture\tablenotemark{a}                       &                  \\
\hline                                                                
\rc(\heplus)              &    pc            &  650   \\
\rx(\heplus)              &    pc            &  1420  \\
F(\hei\ $\lambda 4471$)   &  \flux            & $1.456 \pm 0.004 \times 10^{-15}$    \\
L'(\hei\ $\lambda 4471$)  &  ergs s$^{-1}$   & $ 8.35\pm 0.80 \times 10^{38}$   \\
\enddata
\tablenotetext{a}{\tiny
The symbols \rc(\hplus), \rc(\heplusplus), and \rc(\heplus) denote
the \hplus, \heplusplus, and \heplus\  Str\"{o}mgren radius, respectively.
The symbol \rx\ denotes the extent of the filamentary emission;
see  Fig.~\ref{fig:sb_profiles}.}
%
\tablenotetext{b}{\tiny
Line fluxes given before extinction correction. Line luminosities, L', have
been corrected for extinction. Ionizing photon luminosities $Q(\heplus)$ and $Q(H)$ computed
at $h\nu > 54.4$~eV and $h\nu > 13.6$~eV, respectively.  Recombination rates  are computed at
at the temperature measured in the \oiii\ zone, $T_e \approx 2.0 \times 10^4$~K. 
The ionizing photon efficiency, \xion, is computed from Equation~\ref{eqn:xionH},
using the absolute UV magnitude at 1500 \AA. }
\tablenotetext{c}{\tiny
Electron density, $n_e$, computed from the doublet ratio of the ion indicated;
spectral apertures are identified by name.
Root-mean-square electron density, $\langle n_e^2 \rangle^{1/2}$:  computed from \rc(\hplus) and \Hb\ luminosity
within \rc(\hplus), computed within \rc(\heplusplus)  for the ionization zones probed by 
\heii\ $\lambda 4686$ as well as \Hb.
}
\label{tab:qphot} 
\end{deluxetable*}

\clearpage

\bibliography{ms_eelg}{}
\bibliographystyle{aasjournal}

\end{document}